\documentclass[structabstract]{aa}  
\usepackage{graphicx}
\usepackage{float}
\usepackage{indentfirst}
\usepackage{lscape}
\usepackage{longtable}
\usepackage{txfonts}
\begin{document}
   \title{Modelling the sulphur chemistry evolution in Orion KL}

   \author{G. B. Esplugues
          \inst{1},
          S. Viti  
          \inst{2},  
          J. R. Goicoechea
          \inst{1},
          \and       
          J. Cernicharo
          \inst{1}
         }

   \institute{Centro de Astrobiolog\'ia (CSIC-INTA), Ctra. de Torrej\'on-Ajalvir, km. 4, E-28850 Torrej\'on de Ardoz, Madrid, Spain\\
              \email{espluguesbg@cab.inta-csic.es}
        \and
            Department of Physics \& Astronomy, University College London, Gower St. London WC1E 6BT
             }

   \date{Received ; accepted }

 
  \abstract   
   {We present a study of the sulphur chemistry evolution in the region Orion KL along the gas and grain phases of the cloud. 
   } 
   {Our aim is to investigate the processes that dominate the sulphur chemistry in Orion KL and to determine how physical and chemical parameters, such as the final mass of the star and the initial elemental abundances, influence the evolution of the hot core and of the surrounding outflows and shocked gas (the plateau).}
   {We independently modelled the chemistry evolution of the hot core and the plateau using the time-dependent gas-grain model UCL$\_$CHEM and considering two different phase calculations. Phase I starts with the collapsing cloud and the depletion of atoms and molecules onto grain surfaces. Phase II starts when a central protostar is formed and the evaporation from grains takes place. We show how the stellar mass, the gas density, the gas depletion efficiency, the initial sulphur abundance, the shocked gas temperature, and the different chemical paths on the grains leading to different reservoirs of sulphur on the mantles affect sulphur-bearing molecules at different evolutionary stages for both components. We also compare the predicted column densities with those inferred from observations of the species SO, SO$_2$, CS, OCS, H$_2$S, and H$_2$CS.} 
   {The models that reproduce the observations of the largest number of sulphur-bearing species in both components are those with an initial sulphur abundance of 0.1 times the sulphur solar abundance (0.1S$_{\odot}$) and a density of at least $n$$\mathrm{_{H}}$=5$\times$10$^6$ cm$^{-3}$ in the shocked gas region.}
   {
We conclude that most of the sulphur atoms were ionised during Phase I, consistent with an inhomogeneous and clumpy region where the UV interstellar radiation penetrates and leading to sulphur ionisation. We also conclude that the main sulphur reservoir on the ice mantles was H$_2$S. 
In addition, we deduce that a chemical transition currently takes place in the plateau shocked gas, where SO and SO$_2$ gas-phase formation reactions change from being dominated by O$_2$ to being dominated by OH.}

   \titlerunning{Modelling the sulphur chemistry evolution in Orion KL}
   \authorrunning{G. B. Esplugues et al.}
   \maketitle
%

\section{Introduction}

Sulphur-bearing molecules can be very useful tracers of the chemistry and physical properties of complex star-forming regions (SFRs) located in dense molecular clouds. In particular, they are good tracers of hot cores since they are especially sensitive to physical and chemical variations during the lifetime of these hot and dense regions (e.g. Hatchell et al. 1998; Viti et al. 2004). Hot molecular cores are common in regions of high-mass star formation. They are found in the vicinity of newly formed high-mass stars and are characterised by high temperatures ($\sim$100-300 K) and dense gas ($\sim$10$^{7}$ cm$^{-3}$). Hot cores form during the star formation process and from a chemical point of view are characterised by the evaporation of the grain mantle (formed during the collapse phase). The amount of each atomic or molecular species that evaporates back to the gas phase depends mainly on the binding energies and on the total amount present on the grains at the time the temperature starts increasing. Once back in the gas phase, and under conditions of high temperature and density, the chemistry evolves rapidly, transforming many of these species into more complex ones on timescales of $\lesssim$10$^{5}$ years (Viti 2005). 

According to previous studies (e.g. Hatchell et al. 1998; Viti et al. 2001), the abundances of sulphur-bearing species are strongly time-dependent, and their evolution depends on the heating rate and therefore on the mass of the forming star.
Although the levels of sulphur depletion in different environments are not fully constrained (Goicoechea et al. 2006), OCS and H$_{2}$S ices are probably the main reservoir of sulphur on the grains (Palumbo et al. 1997, Hatchell et al. 1998) before the dust heats up. Druard $\&$ Wakelam (2012) also obtain the same result by analysing the sulphur-depletion efficiency on grains under dense cloud conditions. In particular, they conclude that the lower the (gas and dust) temperature ($<$20 K), the greater the H$_2$S abundance obtained on the grain surfaces. Their study also states that H$_2$S$_2$ and CS$_2$ are other S-bearing molecules efficiently produced on the grains.   
Once these molecules start to sublimate, they rapidly initiate reactions that drive the production of other sulphuretted molecules, such as CS, SO, and SO$_{2}$. Therefore sulphur-bearing species can be used to investigate the evolution of the different phases of the high-mass star formation.

\begin{table*}
\caption{Parameter range used to model the chemistry evolution of the hot core and the plateau of Orion KL.}             
\centering          
\begin{tabular}{c c c c c c c c c c}     
\hline\hline       
          & Size     &  Accretion  & Sulphur      & Maximum           & Star         & Initial gas                     & Final gas                        &   S goes into      &   S$^+$ goes into \\ 
Component & (arcsec) &  efficiency           &abundance   &  temperature $T$ & mass         & density $n$$^{0}_{\mathrm{H}}$ & density $n$$\mathrm{_{H}}$ & mOCS/mH$_2$S       & mOCS/mH$_2$S             \\
          &          &     $f$$_\mathrm{r}$                &(S$_{\odot}$) & (K)               &(M$_{\odot}$) & $\times$10$^2$ (cm$^{-3}$)      & $\times$10$^6$ (cm$^{-3}$)       &  ($\%$)            &  ($\%$)        \\
\hline 

Hot core  & 10       & 0.30-0.85           & 0.01-1       & 300               & 5-15         &   4                                      & 10-100                           & 0-50-100              & 0-50-100       \\
Plateau   & 30       & 0.85              & 0.01-0.1     & 1000-2000         & ...          &   1                                & 0.5-5                             & 0-50-100              & 0-50-100       \\
\hline
\label{table:parameters_grid}                  
\end{tabular}
\tablefoot{Column 9 indicates the percentage of S goes into solid OCS (mOCS) or solid H$_2$S (mH$_2$S) at the end of Phase I, and Col. 10 the percentage of S$^+$ goes into mOCS or mH$_2$S at the end of Phase I.\\
}
\end{table*}

After the gravitational collapse of a molecular cloud, the formation of a new star is accompanied by the development of highly supersonic outflows (Bachiller \& P\'erez-Guti\'errez 1997). The material ejected through these jets collides with the surrounding cloud, compressing and heating the gas, which leads to a drastic alteration of the chemistry because endothermic reactions (or reactions with energy activations barriers) become efficient and dust grains are destroyed. Many observations (Watt et al. 1986, Welch 1988, Bachiller et al. 2001, Esplugues et al. 2013a) have shown that sulphur-bearing species, especially the SO and SO$_{2}$ molecules, show increased column densities (by up to three orders of magnitude) in the shocked regions with respect to more quiescent areas of the cloud. For this reason, these molecules are also considered excellent tracers of shocks.

In this paper we model the sulphur chemistry in the star-forming region Orion KL for two different components, the hot core and the plateau. We present the chemical model in Sect. \ref{chemical model}. For the hot core, we simulate the formation of a star, while in the plateau we simulate the presence of a C-type shock. To study the chemical evolution in both cases, we consider several parameters, such as the star mass and the shock temperature. First, we analyse (Sect. \ref{models_results}) the effects of varying each parameter on the evolution of the SO and SO$_{2}$ abundances. We then compare (Sect. \ref{Comparison}) the model results with observations of different sulphur species (OCS, CS, H$_{2}$CS, SO, SO$_{2}$, and H$_{2}$S). We include a discussion of the results in Sect. \ref{Discussion} and summarize the main conclusions in Sect. \ref{conclusions}.

\section{Chemical model}
\label{chemical model}

To study the evolution of the sulphur chemistry in Orion KL, we used the chemical model UCL$\_$CHEM (Viti 2004). The code is a time-dependent gas-grain model that calculates the chemistry at each time step and/or cloud depth point, providing chemical abundances as a function of time.
The model simulates the chemical evolution of the collapsing cloud, in two-phase calculations: in Phase I (cold phase) the material collapses and atoms and molecules are depleted onto grain surfaces. In this phase, the density increases with time (according to the so-called modified collapse (Spitzer 1978, Nejad et al. 1990), which is described in detail in Rawlings et al. 1992).
In Phase II (warm phase), the density is constant, a central star is formed, and the sublimation from grains takes place due to the warming up of the region. This sublimation can be time-dependent, where mantle species desorb in various temperature bands (Collings et al. 2004), or instantaneous where all species are desorbed off grain surfaces in the first time step. 

Here we employ a time-dependent evaporation, as instantaneous sublimation is a more appropiate approximation only if the mass of the central star(s) is very high ($>$30M$_{\odot}$). The predicted atomic and molecular abundances are sensitive to a wide range of input parameters. The ones we investigate in this paper are the depletion efficiency of accreted species onto grain surfaces, $f$$\mathrm{_r}$, (that is calculated by the fraction of material that is removed from the gas in Phase I); the initial elemental abundances of the main species (H, He, C, O, N, S); the maximum gas temperature; the mass of the central star; the final gas density; and the different paths of the grains leading to different reservoirs of sulphur on the mantles. To choose the range of these input parameters, we considered the descriptions of the Orion-KL physical components found in detailed analyses of the region (Blake et al. 1987, Genzel \& Stutzki 1989, Cernicharo et al. 1994). 
For Phase I, the initial elemental abundances in the gas (relative to the hydrogen nuclei number density and excluding metals in refractory grains) adopted in the chemical code are 1.0, 0.075, 4.45$\times$10$^{-4}$, 1.79$\times$10$^{-4}$, 8.52$\times$10$^{-5}$, and 1.43$\times$10$^{-5}$ for H, He, O, C, N, and S, respectively (Anders \& Grevesse 1989, Asplund 2005, Sofia \& Meyer 2001). We also ran models with depleted sulphur by factors 10 and 100, since it is expected to stick to grains and disappear from the gas phase (see Bergin et al. 2001, Pagani et al. 2005). By the end of Phase I each species will have depleted by different percentages. To define a depletion efficiency, $f$$\mathrm{_r}$, we use the relationship between gaseous and solid CO, since carbon monoxide is the most abundant species after H$_2$, and also since it is now known that its abundance on ices varies among objects (\"Oberg et al. 2011):

\begin{equation}
f_r=mCO/(mCO+CO_{\mathrm{gas}}),
\label{eficiencia}
\end{equation}

\noindent where $mCO$ is the mantle CO at the end of the Phase I of the chemical model. In Phase II the parameter $f$$\mathrm{_r}$ is switched off, because we assume that the radiation from the central star will lead to 100\% efficiency in ice sublimation. 
We have modelled, independently, the chemistry evolution of the hot core and the plateau components of Orion KL. The parameters used to model each region are listed in Table \ref{table:parameters_grid}.
In this version of the code, non-thermal desorption is switched off. For low densities and pre-stellar cores observations, non-thermal desorption mechanisms cannot be ignored. However, given the high densities involved in the hot environments modelled in this paper, and since non-thermal desorption mechanisms and their efficiencies are not determined very well experimentally (e.g. 
see Roberts et al. 2007), adjusting the final percentage of freeze out by reducing the efficiency of the sticking to the grains (via the $f$$\mathrm{_r}$ parameter) is sufficient, therefore avoiding extra free parameters. On the other hand, having Phase I as a separate step calculation allows that the initial molecular fractional abundances used in Phase II calculations are computed by a real time dependence of the chemical evolution of gas-dust interaction processes, i.e. not assumed.

All figures in this paper showing model results are from Phase II. In the model, the times are reset in each phase, so when we show results for Phase II, the time $t$=0 would be when the central star switches on (for hot core models). In the plateau case, the time $t$=0 indicates the start of the shock.

\subsection{Hot core}
\label{Chemical_model_core}

\begin{figure}
   \centering 
   \includegraphics[angle=0,width=9.4cm]{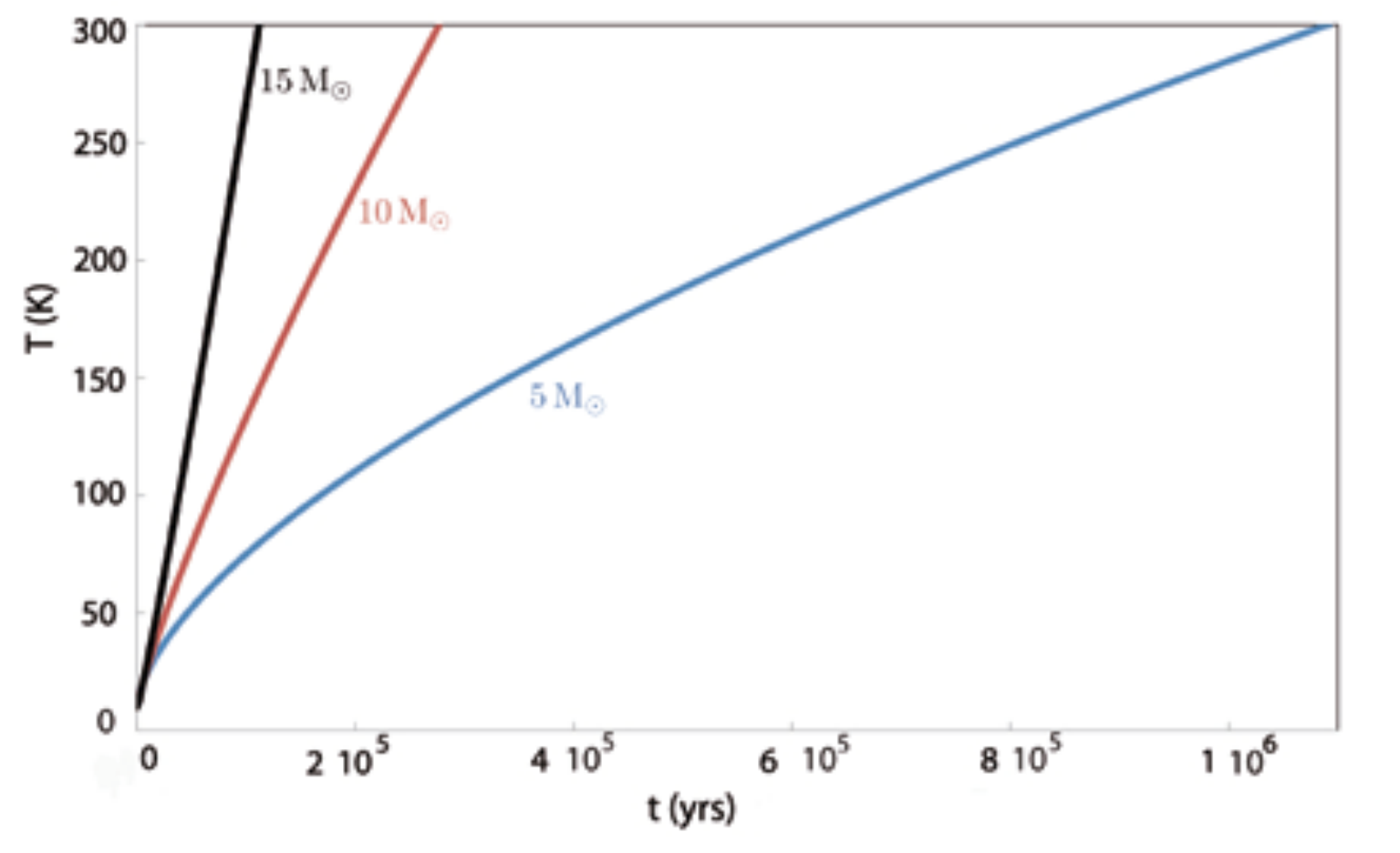}
   \caption{Increase in gas temperature as a time function for three different values of stellar masses (5, 10, and 15M$_{\odot}$). These temperatures are calculated at a distance such that the final visual extinction, Av, is $\sim$400 mag.
}
   \label{figure:tiempos_temperature}
   \end{figure}

To model this region (with an approximated diameter of 10 arcsec), we considered a time-dependent evaporation in Phase II, where the temperature only changes due to the star formed in the central region. In Phase II, we simulated the presence of an infrared source in the center of the core or in its vicinity by increasing the gas and dust temperature up to $T$=300 K, reached at different times depending on the mass of the star formed (Viti et al. 2004). Figure \ref{figure:tiempos_temperature} shows the increase in the gas temperature according to the star mass formed in the center of the core. UCL\_CHEM treatment of the temperature is explained in detail in Viti et al. (2004). Briefly, it is assumed that the temperature of the gas (and dust, as they are coupled at these high densities) is a function of the luminosity (and therefore the age)
of the protostar. Viti et al. (2004) then used the observational luminosity function of
Molinari et al. (2000) to correlate effective temperature, with the age of the
accreting protostar and found that a power law fitted the data. Their Table 2
lists the contraction times (defined as the times after which hydrogen starts burning and the star reaches the zero-age main sequence) as a function of the mass of the star (Viti et al. 2004) and their respective volcano and co-desorption temperatues (which are the temperatues at which the amorphous-to-crystalline H$_2$O ice conversion, the `volcano' effect, and co-desorption when the H$_2$O ice desorbs, respectively).

We note that A$_V$ is calculated as a function of density and size of the core, and it represents the extinction from the irradiating/heating source (not along the line of sight). At the end of Phase I, when the final density is reached, the final A$_V$ is always larger than 400 mags.

\begin{figure}
   \centering 
   \includegraphics[angle=0,width=7.5cm]{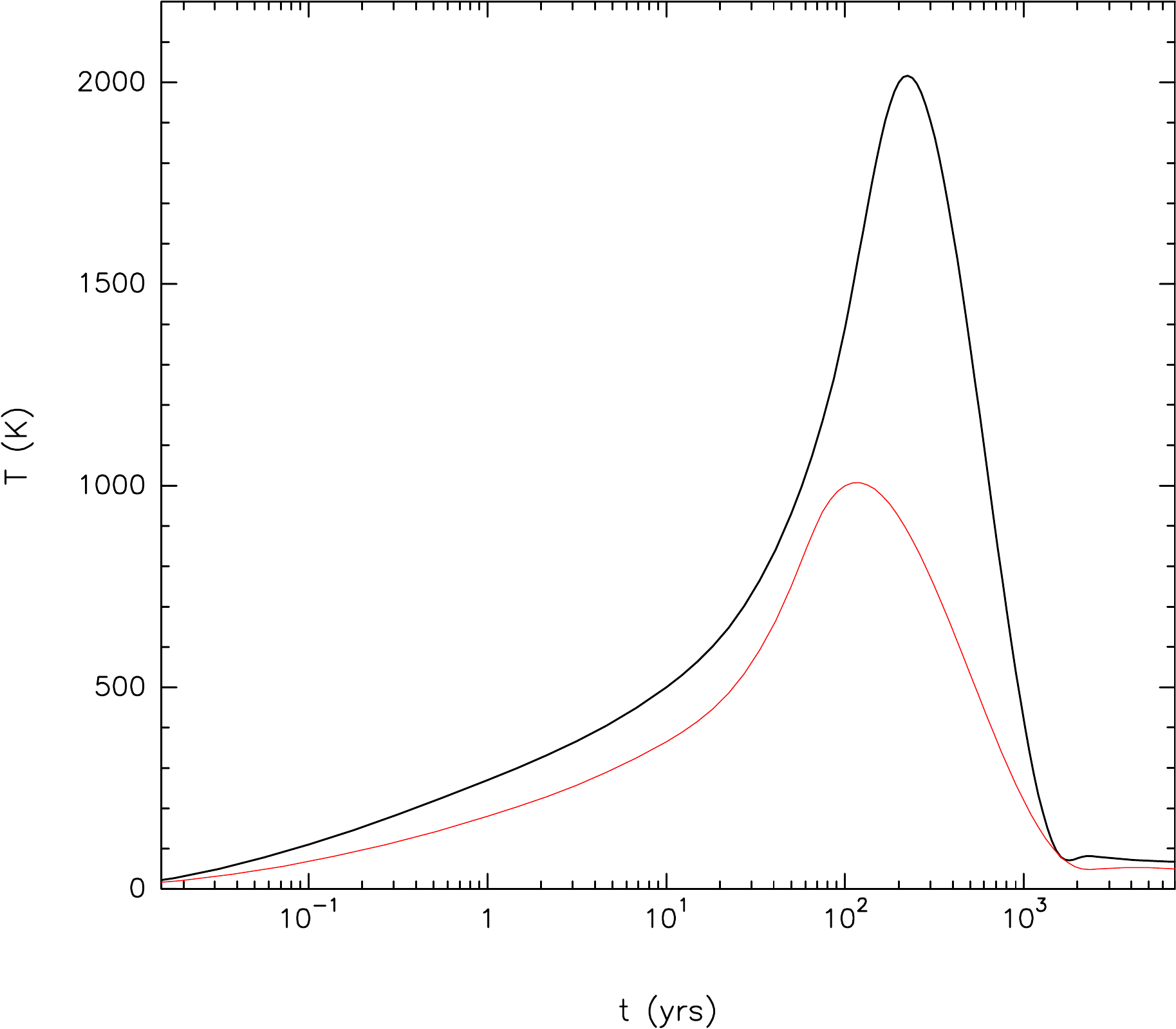}
   \caption{Shocked gas temperatures as a function of time during and after the C-type shock passage ($t$=0 indicates the start of the shock). Black represents the profile where the maximum shocked gas temperature reached is 2000 K and red the case where the maximum temperature is 1000 K. The shock velocities are $\sim$30 and $\sim$20 km s$^{-1}$, respectively. 
}
   \label{figure:shock_temperature}
   \end{figure}

\subsection{Plateau}
\label{Chemical_model_plateau}

The plateau is the region of Orion KL affected by outflows and shocked gas. We use UCL\_CHEM in the same manner as for the hot core by considering a Phase I where the material collapses and a Phase II where $t$=0 corresponds to the start of the shock passage. As in the hot core component, we also assumed time-dependent evaporation in this region. We simulated the presence of a non-dissociative (C-type) shock (Kaufman et al. 1996, Harwit et al. 1998) with a dynamical time scale of $\sim$1000 years (Bally $\&$ Zinnecker 2005), by an increase in the gas temperature. In C-type shocks, the gas is frictionally heated in a layer where the temperatures reached are as high as several thousand kelvin, but the gas remains in molecular form (McKee \& Hollenbach 1980, Neufeld 2001).
We increase the shocked gas temperature up to 1000 K and 2000 K (depending on the model), which correspond to shock velocities of $\sim$20 and $\sim$30 km s$^{-1}$, respectively (Bergin et al. 1998). After the passage of the shock, the gas rapidly (within hundred of years) cools down. The temperature profiles were adopted from the calculations of Bergin et al. (1998), who studied the chemistry of H$_2$O and O$_2$ in post-shock gas. In Fig. \ref{figure:shock_temperature} we plot the temperature evolution in the plateau as a function time during Phase II, where $t$=0 indicates the start of the shock after which a fast cooling of the gas takes place. We considered two different densities,  $n$$\mathrm{_H}$=5$\times$10$^5$ and 5$\times$10$^6$ cm$^{-3}$ in our models (which are the final densities at the end of Phase I), which remain constant along the Phase II of the plateau.

\section{Model results}
\label{models_results}

In this section we describe the influence of the parameters described in Sects. \ref{Chemical_model_core} and \ref{Chemical_model_plateau} on the abundances of SO and SO$_2$. We ran several hot core and plateau models in order to study the sensitivity of the evolution of SO and SO$_2$ abundances to different physical and chemical parameters.

\subsection{Hot core}
\label{hot core}

\begin{figure}
\begin{center}
   \includegraphics[angle=0,width=9cm]{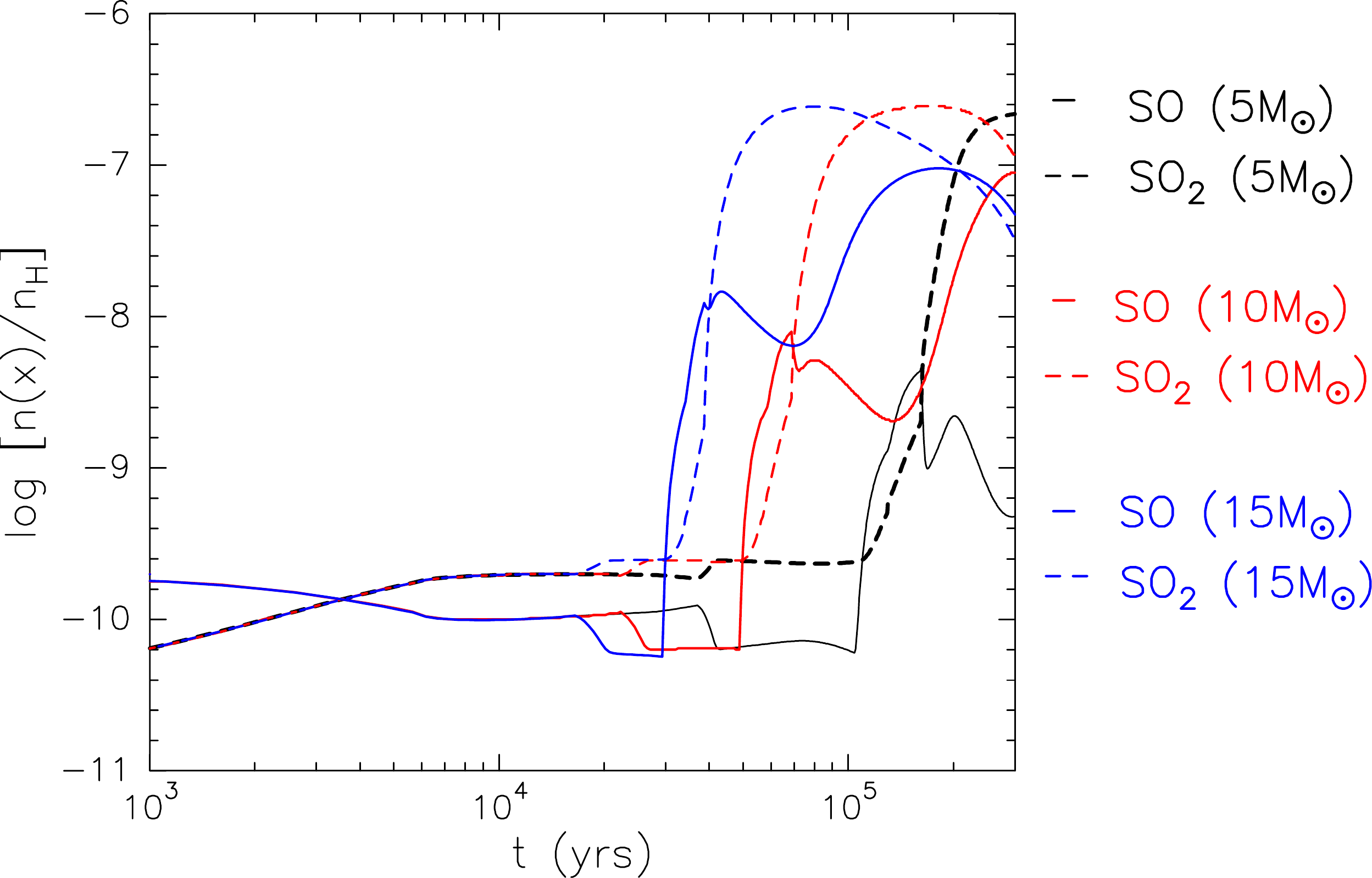} 
   \caption{Differences in the SO and SO$_{2}$ abundance evolution during Phase II for a hot core model when only the star mass is varied. The considered values are 5, 10, and $15\mathrm{M}_{\odot}$ drawn in black, red, and blue, respectively.}
   \label{figure:efecto_masa}
   \end{center}
   \end{figure}

We first analysed the effect of varying the mass of the formed star on the chemical evolution of SO and SO$_2$. Figure \ref{figure:efecto_masa} shows the time evolution of the SO and SO$_{2}$ abundances (in Phase II), for three different types of stars: 5, 10, and 15M$_{\odot}$. For a young ($t$$<$2$\times$10$^{4}$ years) hot core, the abundances of both molecules remain unchanged independently on the mass of the star. 
However, as previous studies have shown (e.g. Viti et al. 2004), the evolutionary stages of both molecules occur at different times strongly depending on the heating rate, i.e., on the mass of the star.
In Fig. \ref{figure:efecto_masa}, we observe a discontinuity in the curves (between 3$\times$10$^4$ and 10$^5$ years, depending on the model) that corresponds to different times at which the sublimation temperature of each species is reached. The difference found is $\sim$5$\times$10$^{4}$ years between a star with 5M$_{\odot}$ and a star with 15M$_{\odot}$.

\begin{figure}
\begin{center}
   \includegraphics[angle=0,width=9cm]{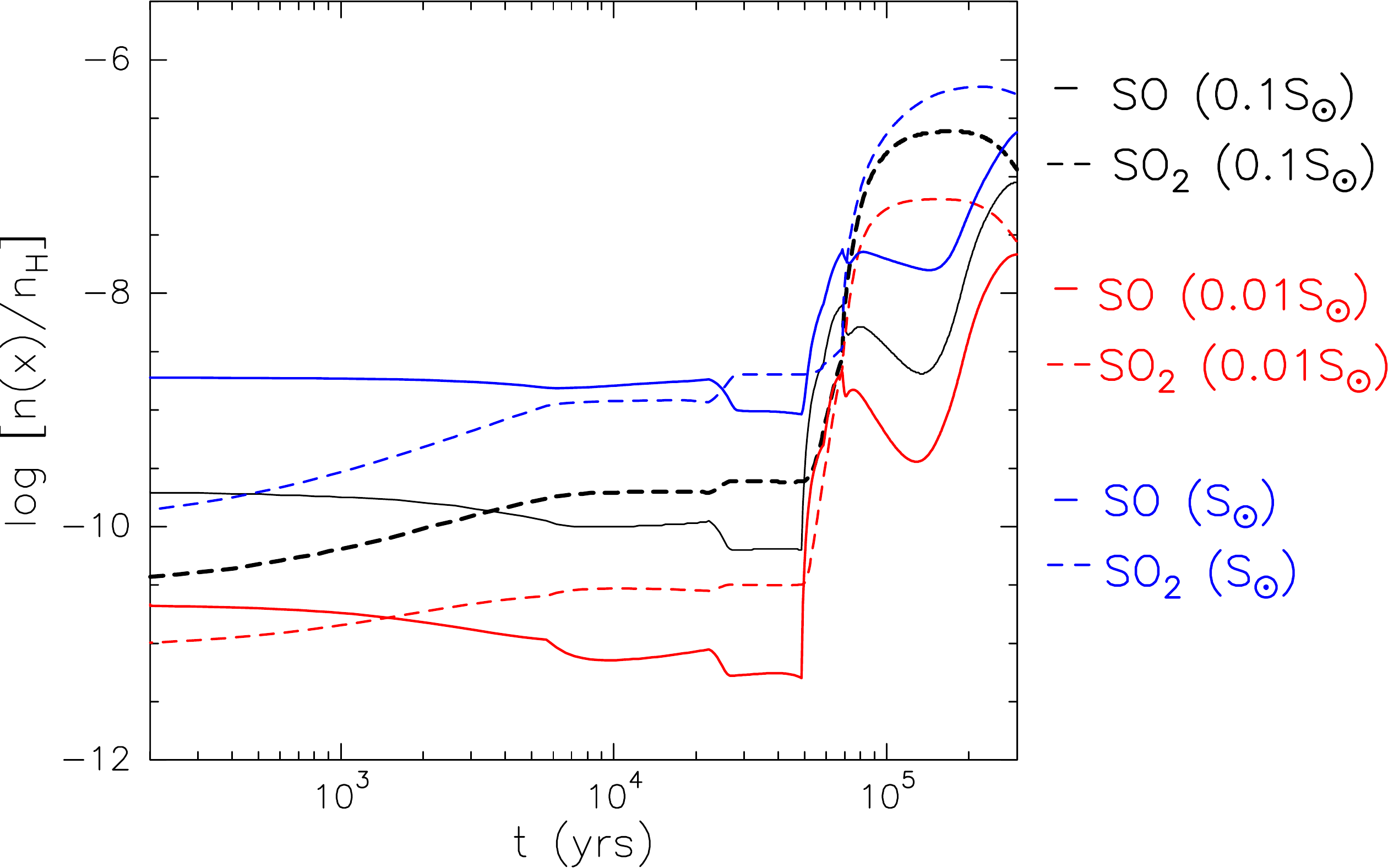} 
   \caption{Differences in the SO and SO$_{2}$ abundance evolution during Phase II for a hot core model when only the initial sulphur abundance is varied. The considered values are 1, 0.1, and 0.01S$_{\odot}$ drawn in blue, black, and red, respectively.}
   \label{figure:efecto_abundancia_azufre}
   \end{center}
   \end{figure}

Figure \ref{figure:efecto_abundancia_azufre} shows how initial sulphur abundance affects the evolution of SO and SO$_{2}$. We follow the abundances of these species as a function of time for three different values of initial sulphur abundance (0.01, 0.1, and 1S$_{\odot}$). For a young hot core, the increase of one order of magnitude in the initial sulphur abundance also leads to the increase of one order of magnitude in the SO abundances; however, this effect is lower as the hot core evolves. In general the SO/SO$_{2}$ ratio increases as the initial sulphur abundance increases.

\begin{figure}
\begin{center}
   \includegraphics[angle=0,width=9.0cm]{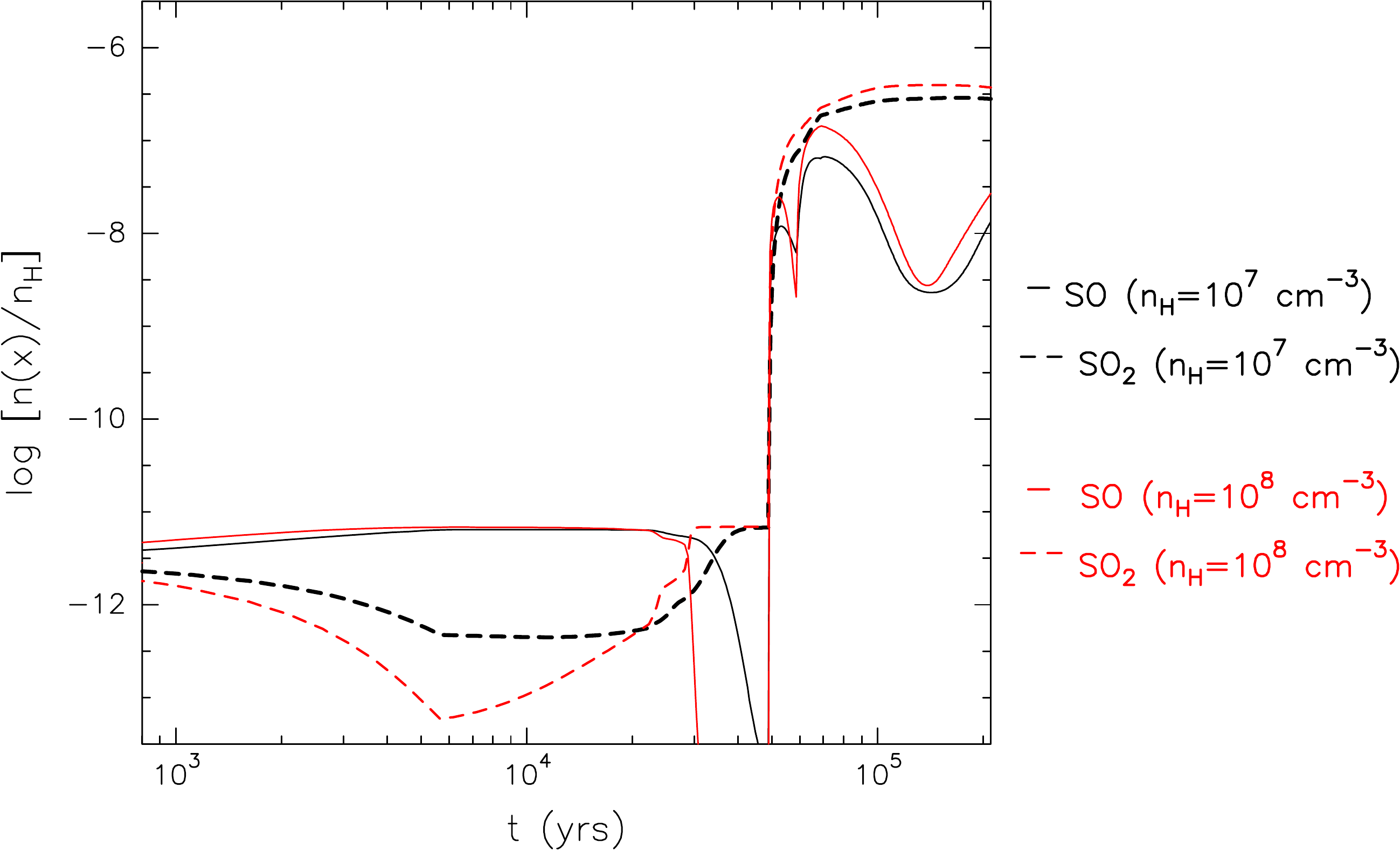} 
   \caption{Differences in the SO and SO$_{2}$ abundance evolution during Phase II for a hot core model when only the final hydrogen density is varied. The considered values are 10$^{7}$ cm$^{-3}$ (black) and 10$^{8}$ cm$^{-3}$ (red).}
   \label{figure:efecto_abundancia_hidrogeno_hot_core}
   \end{center}
   \end{figure}

The effects of varying the hydrogen density on the evolution of SO and SO$_{2}$ abundances during Phase II are shown in Fig. \ref{figure:efecto_abundancia_hidrogeno_hot_core}. We observe the largest differences during the early stage ($t$$<$5$\times$10$^4$), where the increase of one order of magnitude in the hydrogen density of the region mainly provides differences of up to one order of magnitude in the abundances of SO$_2$. For an evolved hot core, the differences produced in the abundances of both species are much lower than one order of magnitude.

\begin{figure}
\begin{center}
   \includegraphics[angle=0,width=9.3cm]{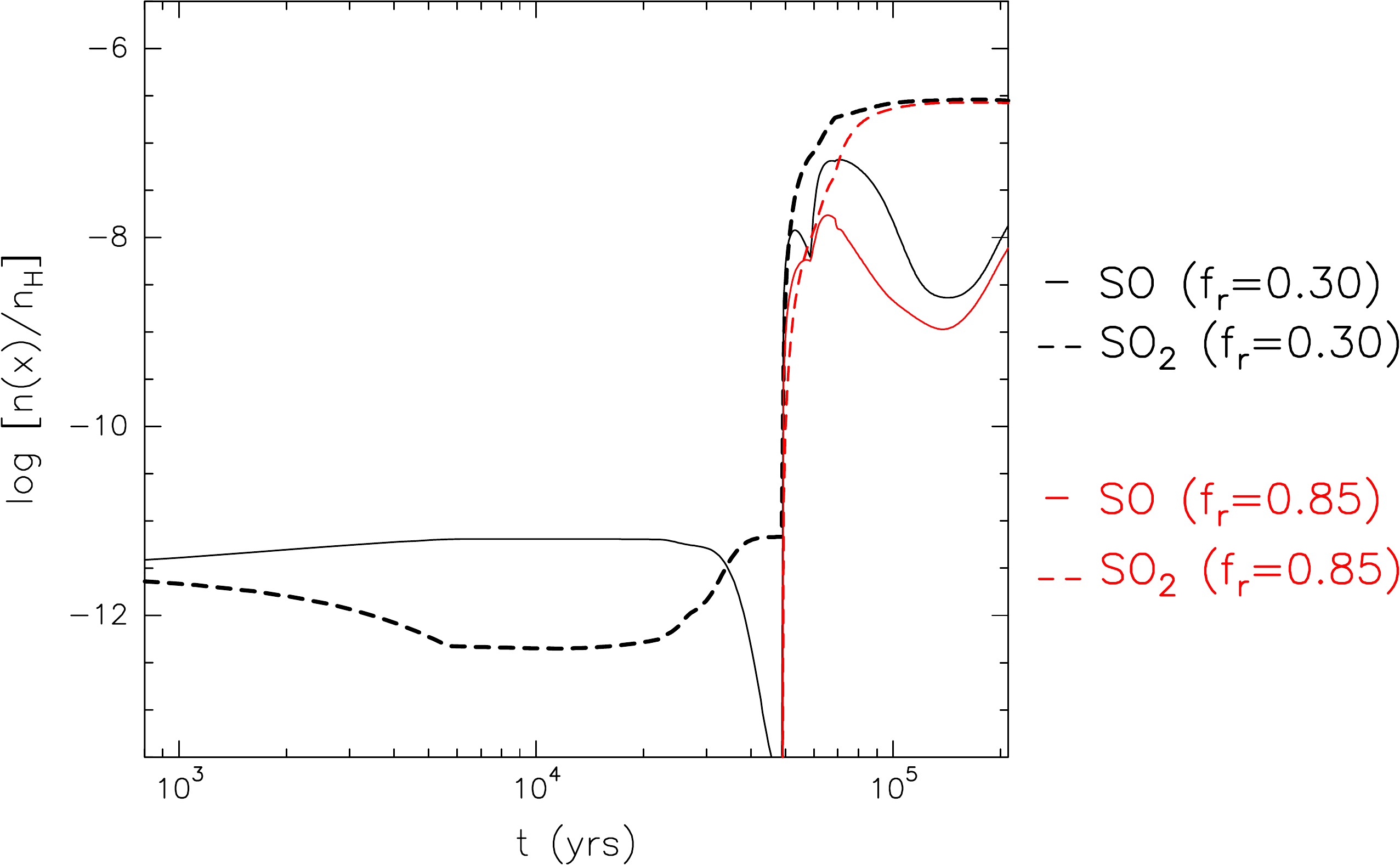} 
   \caption{Differences in the SO and SO$_{2}$ abundance evolution during Phase II for a hot core model when only the accretion efficiency is varied. The considered values are 0.30 in black and 0.85 in red.}
   \label{figure:efecto_eficiencia_hot_core}
   \end{center}
   \end{figure}

Figure \ref{figure:efecto_eficiencia_hot_core} shows the evolution of SO and SO$_{2}$ during the warm phase (Phase II) of the hot core as a function of the gas accretion efficiency, $f$$_{\mathrm{r}}$, occurred during Phase I. We have considered two values of $f$$_{\mathrm{r}}$; 0.3 and 0.85, which imply CO depletions of $\sim$1$\%$ and $\sim$99$\%$, respectively, obtained through the Eq. \ref{eficiencia}.  
We deduce that the accretion efficiency during Phase I is especially relevant for the SO and SO$_2$ abundances in Phase II. In particular, for a young hot core ($t$$<$5$\times$10$^{4}$ years), differences of more than two orders of magnitude in the abundances of both species are found depending on whether the accretion efficiency is low (0.3) or high (0.85). 
For longer times ($t$$\gtrsim$10$^{5}$ years), the effects of varying this parameter become almost negligible, especially for SO$_2$.

\begin{figure}
\begin{center}
   \includegraphics[angle=0,width=9.3cm]{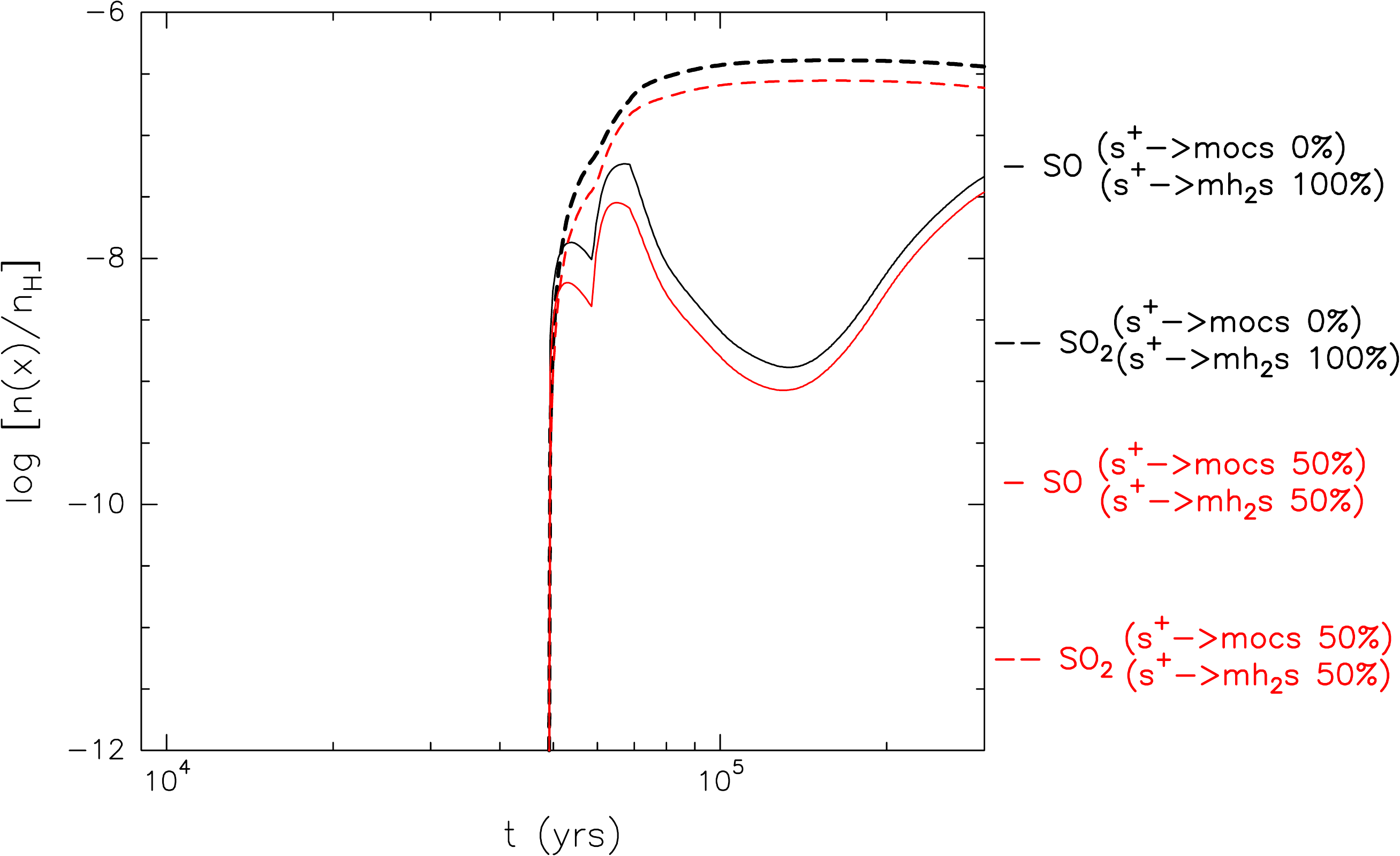} 
   \caption{Effect on SO and SO$_{2}$ abundances (Phase II) of varying the percentage of S$^{+}$ frozen into H$_{2}$S and OCS. In black, we show the results when all S$^+$ freezes out to form H$_{2}$S and in red, when S$^+$ freezes out to form OCS (50\%) and H$_{2}$S (50\%) at the end of Phase I.}
   \label{figure:efecto_porcentaje_S+_sobre_SO-SO2}
   \end{center}
   \end{figure} 

\begin{figure}
\begin{center}
   \includegraphics[angle=0,width=9.3cm]{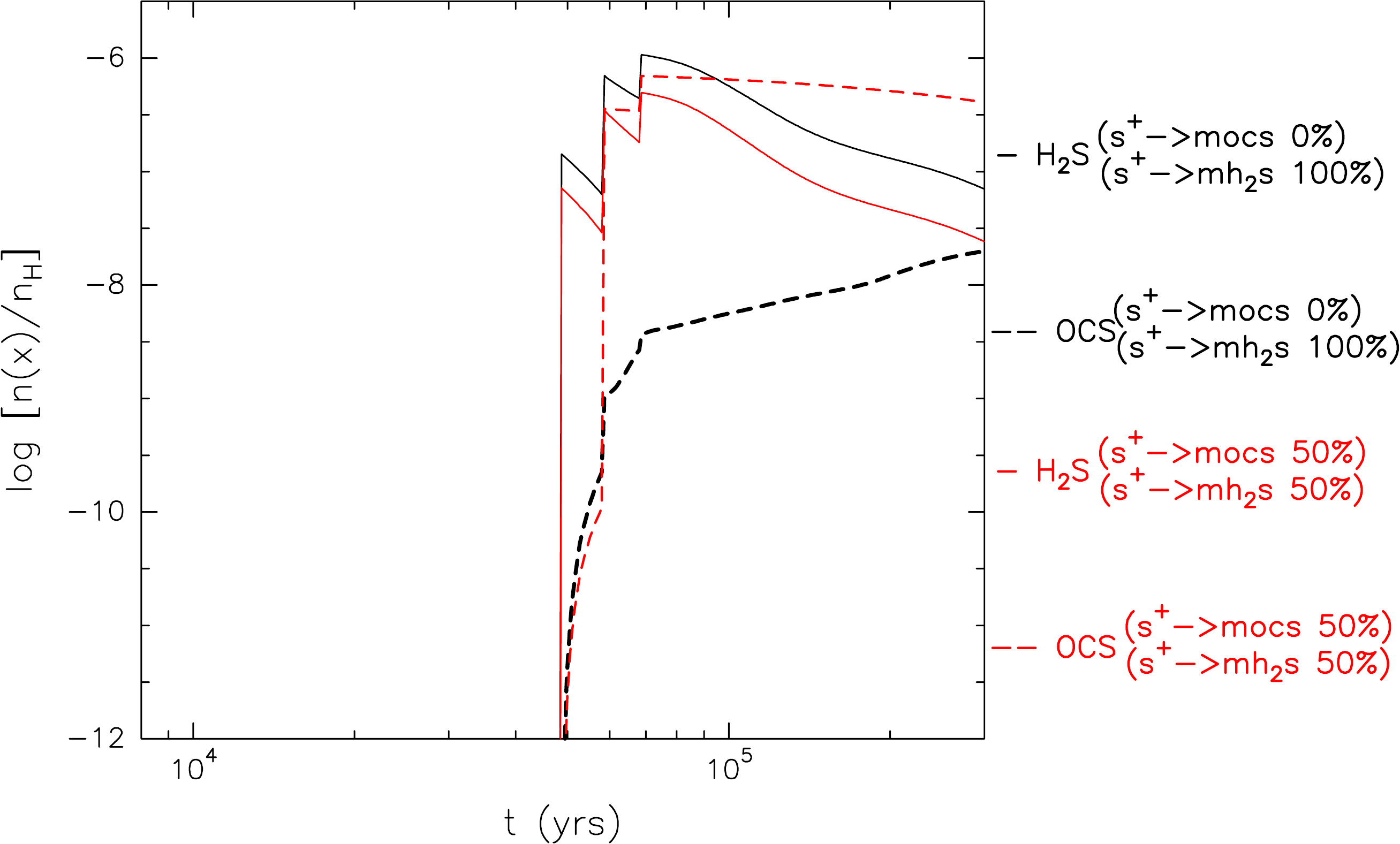} 
   \caption{Effect on OCS and H$_{2}$S abundances (Phase II) of varying the percentage of S$^{+}$ frozen into H$_{2}$S and OCS. In black, we show the results when all S$^+$ freezes out to form H$_{2}$S and in red, when S$^+$ freezes out to form OCS (50\%) and H$_{2}$S (50\%) at the end of Phase I.}
   \label{figure:efecto_porcentaje_S+}
   \end{center}
   \end{figure}   

We have also looked at how abundances of SO and SO$_2$ in Phase II are affected by the form sulphur takes once on the grains. Atoms, such as sulphur in gas phase accreting on the grains from the gas, lead to the formation of new species in the solid phase. 
Given that H is the most abundant element, we assume that the hydrogenation of S atoms is energetically viable; S atoms will tend to form H$_2$S molecules because they impinge on icy grain mantles. 
We have investigated the effects of varying the percentage of sulphur atoms (S and S$^+$) that freeze out to form H$_{2}$S and OCS on the mantles at the end of Phase I. We find that varying the amount of neutral S that goes into H$_2$S or OCS does not significantly affect the abundances of SO and SO$_2$ during Phase II. We also find only a very slight change in the abundances of OCS. For the ionised case, when we consider that all S$^{+}$ atoms freeze out to form H$_{2}$S, we obtain similar results to those obtained for the case where different percentages of neutral sulphur freeze out to form OCS and H$_2$S. After including in the model the possibility that part of S$^{+}$ also goes into OCS on the mantles (Figs. \ref{figure:efecto_porcentaje_S+_sobre_SO-SO2} and \ref{figure:efecto_porcentaje_S+}), we obtain small differences (lower than one order of magnitude) among the abundances of SO, SO$_2$, and H$_2$S. The most important difference is found in the OCS evolution, where its abundance varies up to two orders of magnitude in Phase II, when only the 50\% of S$^{+}$ is frozen out to form H$_{2}$S at the end of the Phase I. All together, this suggests that most of the sulphur (whose ionisation potential is relatively low 10.36 eV) present in the diffuse phase of the cloud (Phase I) must be ionised and that it is frozen out before being neutralised. This agrees with Ruffle et al. (1999), who proposed that S$^+$ would freeze out onto dust grains during the collapse more efficiently than neutral species, given that in dense interstellar regions, grains typically carry one negative charge (Gail \& Sedlmayr 1975, Bel et al. 1989, Druard \& Wakelam 2012). The time-scale of collision of S$^+$ with grains on one order of magnitude smaller than those associated with the neutral species containing most of the carbon, oxygen, and nitrogen. This result is also consistent with the detection of S recombination lines in dark clouds (Pankonin \& Walmsley 1978).

\subsection{Plateau}
\label{results_plateau}

\begin{figure}[h!]
\begin{center}
\includegraphics[angle=0,width=9cm]{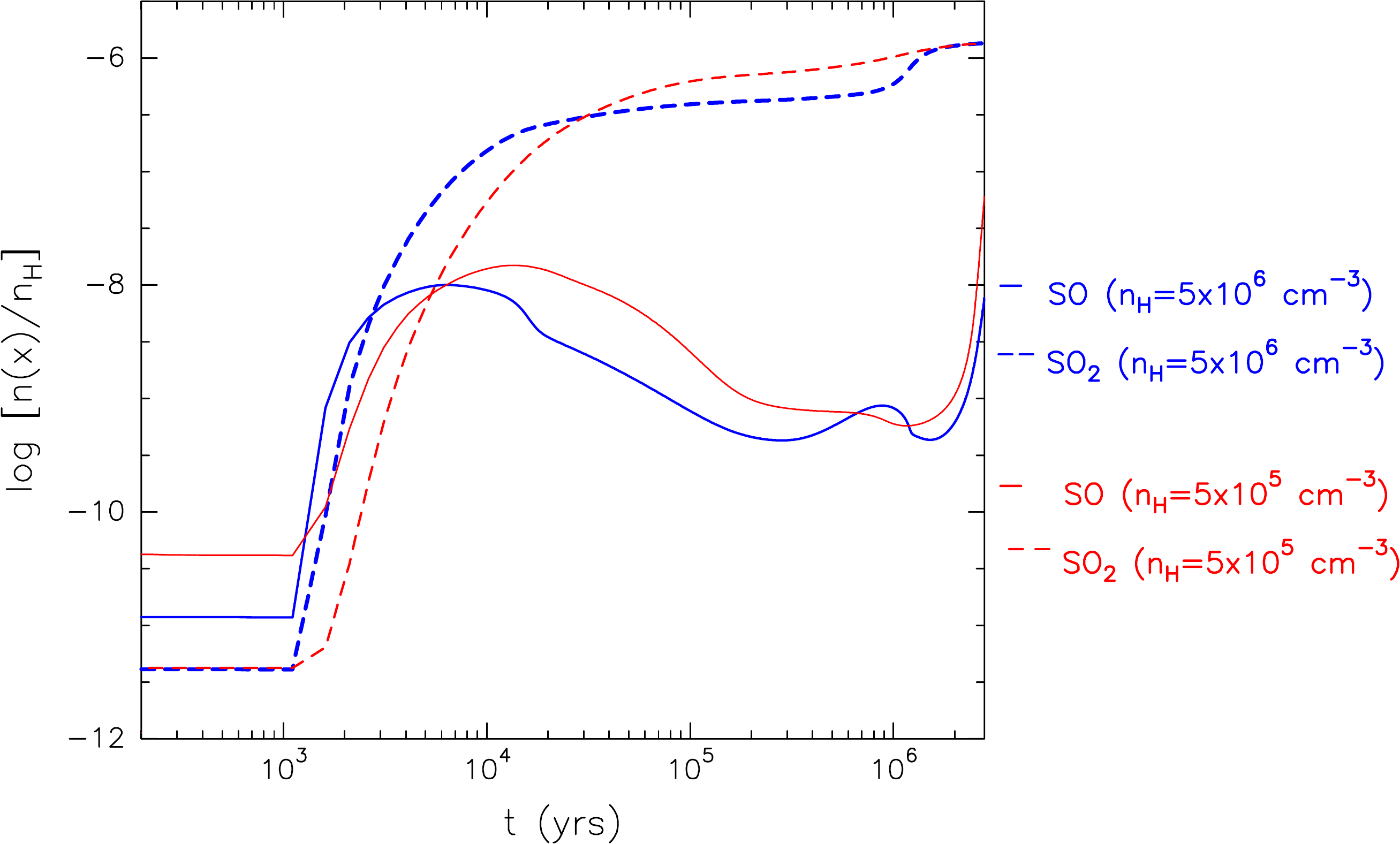} 
   \caption{Differences in the SO and SO$_{2}$ abundance evolution during Phase II for a plateau model when only the final gas density is varied. The considered values are 5$\times$10$^{6}$ cm$^{-3}$ (blue) and 5$\times$10$^{5}$ cm$^{-3}$ (red). 
   }
   \label{figure:efecto_abundancia_hidrogeno_plateau}
   \end{center}
   \end{figure}

Figure \ref{figure:efecto_abundancia_hidrogeno_plateau} shows the evolution of the abundance of SO and SO$_2$ during Phase II when different values of gas density are considered. Variations of one order of magnitude in the gas density can lead to differences of up to two orders of magnitude in the abundances of these species. In particular, we observe this trend for the early stage of the plateau, especially during the gas cooling after the shock (4$\times$10$^2$$\lesssim$$t$$\lesssim$2$\times$10$^3$ yrs). For an evolved plateau, the differences obtained in the abundances of SO and SO$_2$ are less than one order of magnitude, becoming negligible for $t$$>$10$^6$ years.

\begin{figure}[h!]
\begin{center}
   \includegraphics[angle=0,width=9cm]{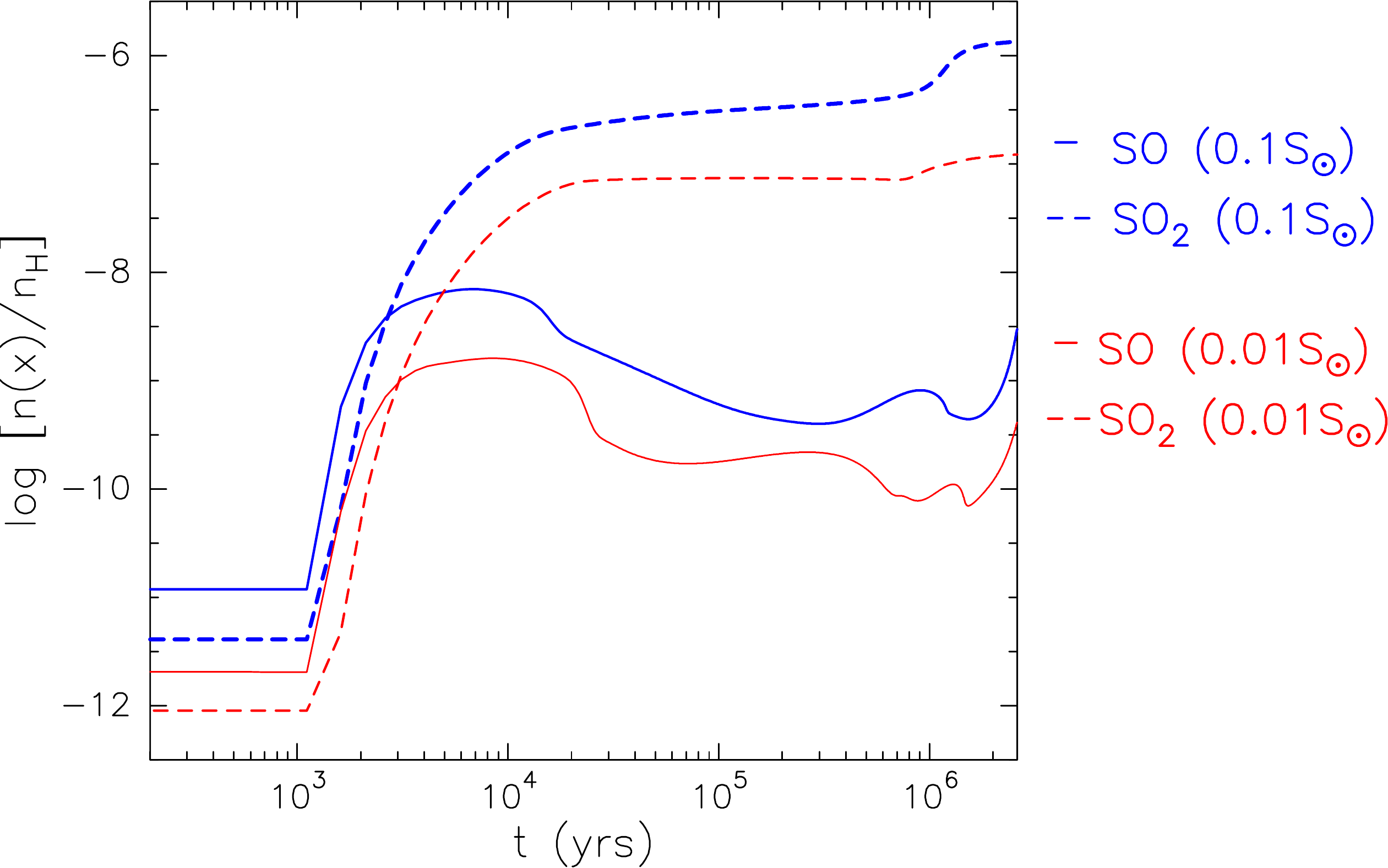} 
   \caption{Differences in the SO and SO$_{2}$ abundance evolution during Phase II for a plateau model when only the initial sulphur abundance is varied. The considered values are 0.1S$_{\odot}$ (blue) and 0.01S$_{\odot}$ (red).}
   \label{figure:efecto_abundancia_solar_plateau}
   \end{center}
   \end{figure}
   
\begin{figure}[h!]
\begin{center}
   \includegraphics[angle=0,width=9.0cm]{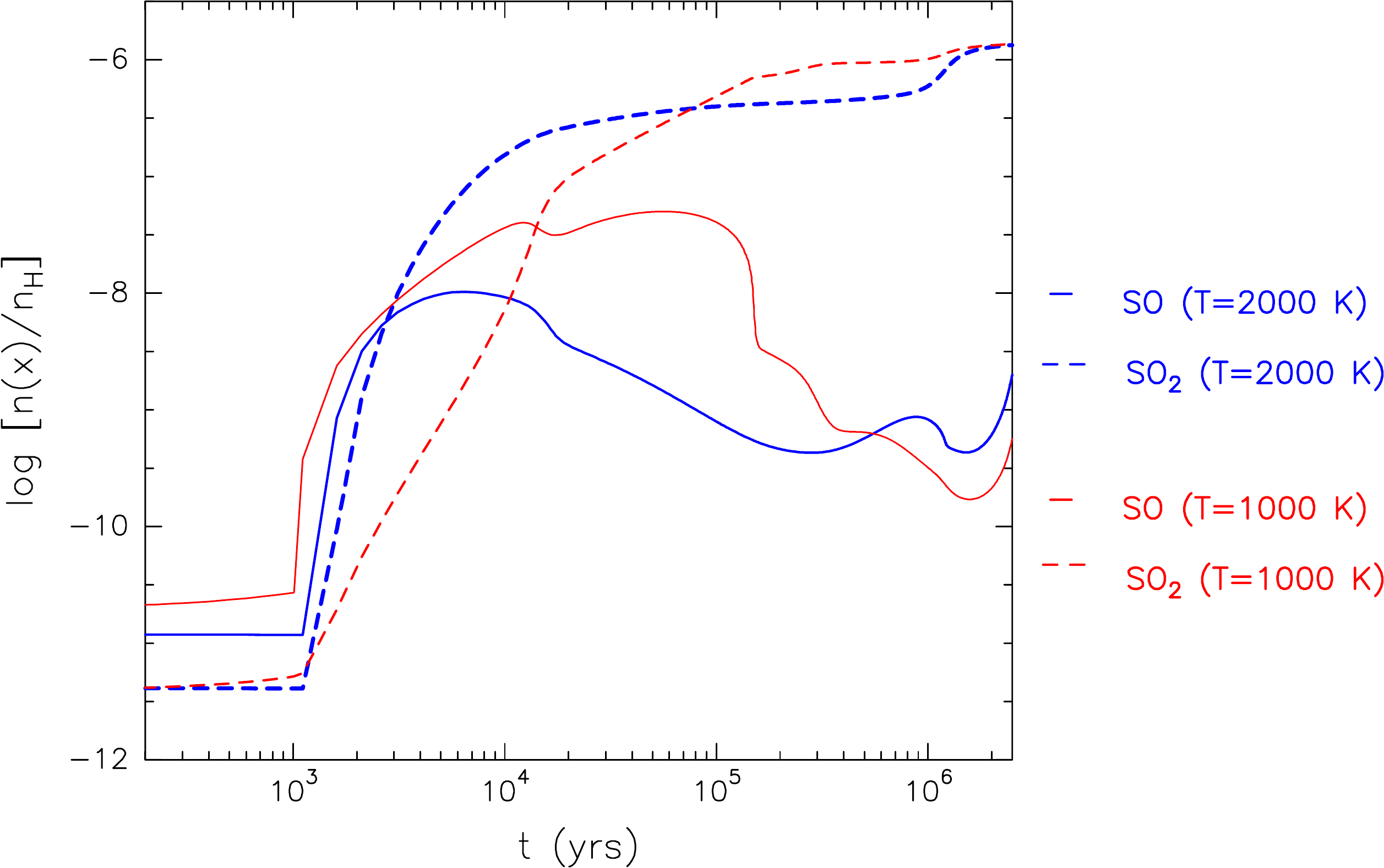} 
   \caption{Differences in the SO and SO$_{2}$ abundance evolution during Phase II for a plateau model when only the maximum shocked gas temperature is varied. The considered values are 2000 K (blue) and 1000 K (red).}
   \label{figure:efecto_Tshock_plateau}
   \end{center}
   \end{figure}

We have also analysed the effect of varying the initial sulphur abundance on the SO and SO$_{2}$ evolution in the plateau. We considered two values (0.1 and 0.01S$_{\odot}$). In Fig. \ref{figure:efecto_abundancia_solar_plateau} we show that the increase of one order of magnitude in the initial sulphur abundance leads to the obvious increase of up to one order of magnitude in the abundances of SO and SO$_{2}$ during the warm phase. Only during the gas cooling the differences found in the abundances are slightly larger than one order of magnitude.

Figure \ref{figure:efecto_Tshock_plateau} shows the results of varying the gas temperature due to the presence of C-shock on the SO and SO$_{2}$ evolution during Phase II. We have considered two possible values for the maximum temperature reached by the gas during the shock, 1000 K and 2000 K. During the fast increase in the gas temperature (see Fig. \ref{figure:shock_temperature} with the temperature profiles as a time function), we find only small variations in the SO$_{2}$ abundances, while the differences found for SO are slightly larger. In particular, since the start of the shock passage (at $t$=0) and up to $t$$\sim$10$^5$ years, we predict that a higher maximum gas temperature reached leads to larger SO$_2$ abundances and lower SO abundances.

\section{Comparison with observations}
\label{Comparison}

To compare model results and observations, we ran several hot core and plateau models considering the values for the different input parameters used in Sect. \ref{models_results}. From the obtained abundances of each species, we estimated their column densities, $N_x$,

\begin{equation}
\label{conversor_abundances}
N_x = (n_x/n_{\mathrm{H}})\ A_V\,1.6\times10^{21},
\end{equation}

\noindent where ($n_x$/$n_{\mathrm{H}}$) is the abundance of each species, $A_V$ the visual extinction, and 1.6$\times$10$^{21}$ cm$^{-2}$ the hydrogen column density at 1 mag of extinction. 
We compared these results with the column densities of SO, SO$_{2}$, CS, OCS, H$_{2}$CS, and H$_{2}$S (see Table \ref{table:column_densities}) inferred from the IRAM 30m and Herschel/HIFI line surveys (Tercero et al. 2010, Crockett et al. 2014b, respectively).

These source-averaged column densities were obtained by the same method, by fitting a large number of rotational lines using a non-local thermodynamic equilibrium (non-LTE) excitation and radiative transfer code, MADEX\footnote{The emission observed of the sulphur-bearing molecules considered in this paper was analysed using the code MADEX, except for H$_2$S, whose emission was analysed using the radiative transfer code RADEX (van der Tak et al. 2007). See more details in Crockett et al. (2014a).} (Cernicharo 2012), where it is assumed that the width of the lines is due to the existence of large velocity gradients (LVGs) across the cloud. The radiative coupling between two relatively close points is thus negligible, and the excitation problem is local. Only for the hot core, we considered LTE excitation, which means that most transitions are thermalized to the same temperature. Corrections for beam dilution were also applied for each line depending on the different beam sizes at different frequencies.
For the hot core and the plateau of Orion KL, uniform physical conditions of density, kinetic temperature, line width, and radial velocity ($v$$_{\mathrm{LSR}}$) were assumed taking the results obtained from Gaussians fits of the line profiles, rotational diagrams, and different maps of each region into account (see Tercero et al. 2010 and Crockett et al. 2014b). In addition, in the case of the density, typical values quoted in the literature were also considered. Table \ref{table:components_MADEX} lists the values assumed for each parameter, where only the column density of each component was left as a free parameter in MADEX. 

To determine the uncertainty of the values of hydrogen density and of the kinetic temperature, several MADEX models were run, varying only the values for these parameters and fixing the rest. Comparing the intensity differences between the spectra and the obtained line profiles, uncertainties of 20$\%$ and 15$\%$, were deduced for the temperature and the hydrogen density, respectively. 
The column density of each species was taken from the model that best reproduced the majority of the observed line profiles from transitions covering a wide energy range within a 20$\%$ of the uncertainty in the line intensity. To determine the column density errors shown in Table \ref{table:column_densities} we took different sources of uncertainty into account, such as spatial overlap of the cloud spectral components, pointing errors during the observations, and the lack of collisional rates for some species (SO$_2$, H$_2$CS).

With respect to the model results, the main uncertainties are in the treatment of the surface reactions, in the assumed initial elemental abundances in the gas (listed in Sect. \ref{chemical model}), with uncertainties up to 10$\%$ (from a comparison with other works), and in the rates for the chemical reactions available in gas-phase networks (e.g. Wakelam et al. 2012).

\begin{table}
\caption{Source-averaged column densities, $N$, inferred from observations towards Orion KL.}             
\centering   
\begin{tabular}{c c c c}     
\hline\hline       
         & Plateau              & Hot core              &              \\ 
Species  & $N$$\times$10$^{15}$ & $N$$\times$10$^{15}$  & References   \\ 
         & (cm$^{-2}$)          & (cm$^{-2}$)           &              \\ 
\hline                    
SO       &  18$\pm$10           & 9$\pm$4               &   (1)        \\  
SO$_{2}$ &  10$\pm$4            & 65$\pm$20             &   (1)        \\
CS       &  2.4$\pm$0.5         & 1.7$\pm$0.9           &   (1, 2)     \\
OCS      &  12$\pm$3            & 15$\pm$4              &   (2)        \\
H$_{2}$CS&  1.0$\pm$0.3         & 1.6$\pm$0.5           &   (2)        \\
H$_{2}$S &  ...                 & 950$\pm$200           &   (3)        \\
\hline 
\label{table:column_densities}                 
\end{tabular}
\tablebib{
(1) Crockettt et al. (2014b); (2) Tercero et al. (2010); (3) Crockett et al. (2014a).}\\
\end{table}

\begin{table}
\caption{Parameters adopted for the hot core and the plateau of Orion KL in MADEX.}             
\centering   
\begin{tabular}{c c c c c }     
\hline\hline       
Component & Density & Temperature  & $\triangle$$\nu$$_{\mathrm{FWHM}}$ & v$_{\mathrm{LSR}}$ \\
          & (cm$^{-3}$)                     & (K) & (km s$^{-1}$) & (km s$^{-1}$) \\  
\hline                    
   Plateau   & 1$\times$10$^{6}$ & 125 & 25 & 6 \\
   Hot core  & 5$\times$10$^{7}$ & 225 & 10 & 5.5 \\
\hline
\label{table:components_MADEX}                  
\end{tabular}
\end{table}

\subsection{Hot core}
\label{comparison_hot_core}

We list in Table \ref{table:hotcore_model} the hot core models run considering the input parameters described in Sect. \ref{Chemical_model_core}. In figures for Sects. \ref{comparison_hot_core} and \ref{comparison_plateau}, we show (with points\footnote{Since the observations do not have measured times, their representation in Figs. of Sects. \ref{comparison_hot_core} and \ref{comparison_plateau} should be through horizontal lines. We have only plotted points (on which horizontal lines would be located) to indicate the time at which model results are matched to the observations well.}) when model results reproduce the observational results of the species SO, SO$_2$, CS, OCS, H$_2$CS, and H$_2$S.

\begin{table}
\caption{Hot core chemical models.}             
\centering          
\begin{tabular}{lc c c  c c}     
\hline\hline       
            
           & Star     & Sulphur  & Accretion  & Density  \\ 
Model      & mass &  abundance   & efficiency    &  $n$$\mathrm{_H}$              \\
       & (M$_{\odot}$) &   (S$_{\odot}$)    &  $f$$_\mathrm{r}$   &  $\times$10$^7$ (cm$^{-3}$)               \\
\hline 

   1\textbar2\textbar3    &  15\textbar10\textbar5 & 0.1  & 0.3  & 1-10 \\
   4\textbar5\textbar6    &  15\textbar10\textbar5 & 0.1  & 0.85 & 1-10 \\  
   7\textbar8\textbar9    &  15\textbar10\textbar5 & 0.01 & 0.3  & 1-10  \\
   10\textbar11\textbar12 &  15\textbar10\textbar5 & 0.01 & 0.85 & 1-10 \\
   13\textbar14\textbar15 &  15\textbar10\textbar5 & 1    & 0.3  & 1-10 \\
   16\textbar17\textbar18 &  15\textbar10\textbar5 & 1    & 0.85 & 1-10  \\

\hline
\label{table:hotcore_model}                  
\end{tabular}
\tablefoot{Models with a density of 10$^8$ cm$^{-3}$ are noted by the numbers of Col. 1 and adding the letter \textit {b}.\\
}
\end{table}

For all models with a stellar mass of 5M$_{\odot}$ (see for example Fig. \ref{figure:hotcore_model_18}, Appendix \ref{Figures of Hot core}), the observations are reproduced after $t$$>$10$^{5}$ years since the central star switches on which, if compared to the dynamical timescale, it is possibly too long for the hot core of Orion KL whose age is estimated to be closer to $\sim$10$^{4}$ years (Walmsley et al. 1987, Plambeck \& Wright 1987, Brown et al. 1988). In addition, the time range at which all sulphur-bearing species are matched well to the observations (taking the error bars of the observed column densities into account) is up to $\sim$5$\times$10$^4$ years, which is probably too long a time.

With models where an initial solar sulphur abundance has been considered (see for example Model 17 in Fig. \ref{figure:hotcore_model_17}), it is also possible to reproduce the observations of the sulphur-bearing species; however this value leads to column densities of H$_2$S too high (see also Fig. \ref{figure:hotcore_model_18}). Therefore we focus on models with initial sulphur abundances of 0.1S$_{\odot}$ and 0.01S$_{\odot}$.  

Figures \ref{figure:hotcore_model_11} and \ref{figure:hotcore_model_10} (Models 11 and 10 with 10M$_{\odot}$ and 15M$_{\odot}$, respectively) show the run models considering an initial sulphur abundance of 0.01S$_{\odot}$, $f$$\mathrm{_r}$=0.85, and $n$$\mathrm{_H}$=10$^7$ cm$^{-3}$. In both cases we notice that the observations of OCS, H$_2$S, and SO are not reproduced, even taking the error bars into account, since the maximum column densities obtained with these models are $\sim$1 order of magnitude lower than those observed. SO$_2$ is not reproduced either, but the difference between observations and models is less than one order of magnitude. We have also run Models 11 and 10 with a gas density of 10$^8$ cm$^{-3}$ instead of 10$^7$ cm$^{-3}$ (Figs. \ref{figure:hotcore_model_11b} and \ref{figure:hotcore_model_10b} with Models 11\textit{b} and 10\textit{b}, respectively): in this case, it is possible to reproduce the observations of the sulphur-bearing species.

\begin{figure}
\begin{center}
   \includegraphics[angle=0,width=7.5cm]{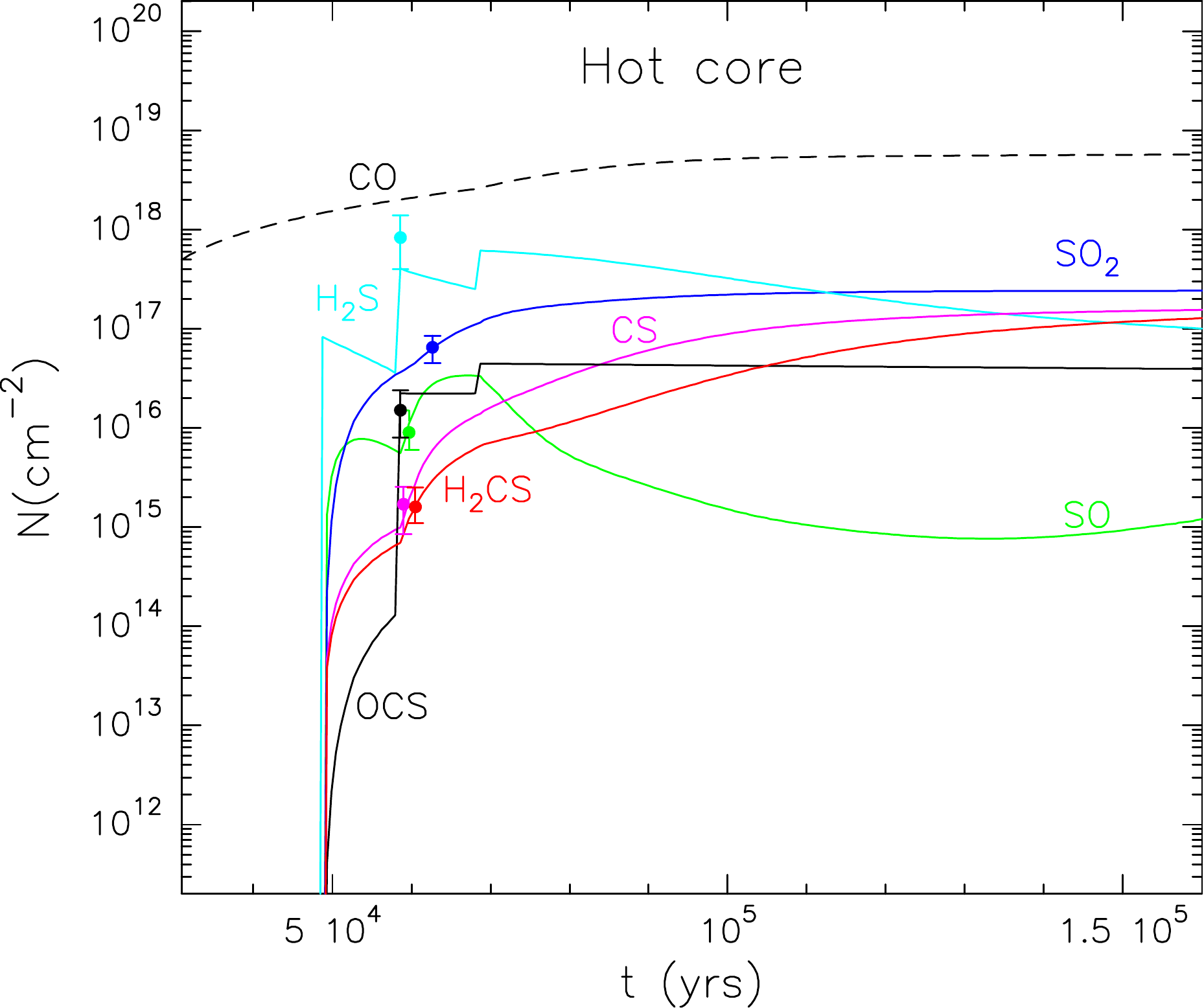} 
   \caption{Sulphur-bearing species column densities as a time function (gas phase) for the hot core Model 5 with a star mass of 10M$_{\odot}$, $f$$\mathrm{_r}$=0.85, a 0.1S$_{\odot}$ initial sulphur abundance, and a density of 10$^7$ cm$^{-3}$. The points indicate the observational results of SO, SO$_{2}$, CS, OCS, H$_{2}$CS, and H$_{2}$S.}
   \label{figure:hotcore_model_5}
   \end{center}
   \end{figure}

Model 5 is shown in Fig.\ref{figure:hotcore_model_5}, where a stellar mass of 10M$_{\odot}$, a hydrogen density of 10$^7$ cm$^{-3}$, and an intermediate initial sulphur abundance of 0.1S$_{\odot}$ have been assumed. In this case, the observations of the six sulphur-bearing species (SO, SO$_{2}$, CS, OCS, H$_{2}$CS, and H$_{2}$S) are reproduced at $t$$\sim$6$\times$10$^{4}$ years after the central star switches on, independently of the considered accretion efficiency (see Model 2, Fig. \ref{figure:hotcore_model_2}, with $f$$\mathrm{_r}$=0.3). 
Fig. \ref{figure:hotcore_model_4} shows that Model 4 (similar to Model 5, but with a central star of 15M$_{\odot}$) also provides a good fit.

All together these results suggest that models with an initial sulphur abundance of 0.1S$_{\odot}$ and a hydrogen density of 10$^7$ cm$^{-3}$, as well as models assuming 0.01S$_{\odot}$ and  $n$$\mathrm{_H}$=10$^8$ cm$^{-3}$, with a central star of 10-15M$_{\odot}$ in both cases, can reproduce the observations of the molecules SO, SO$_2$, CS, OCS, H$_2$, and H$_2$CS in the hot core of Orion KL.  
In Esplugues et al. (2013b), we also modelled the species HC$_3$N and DC$_3$N in the hot core of Orion KL. We only reproduced the observations running a model with $n$$\mathrm{_H}$=10$^7$ cm$^{-3}$, a high depletion efficiency (85$\%$), and a central star of 10M$_{\odot}$ (because too fast an increase in temperature, as occurs for 15M$_{\odot}$, leads to a more efficient destruction chemistry, hence misses the enhancement phase for HC$_3$N and DC$_3$N). This therefore favours Model 5 (Fig. \ref{figure:hotcore_model_5}) in our present grid.

\begin{figure}
   \centering 
   \includegraphics[angle=0,width=8cm]{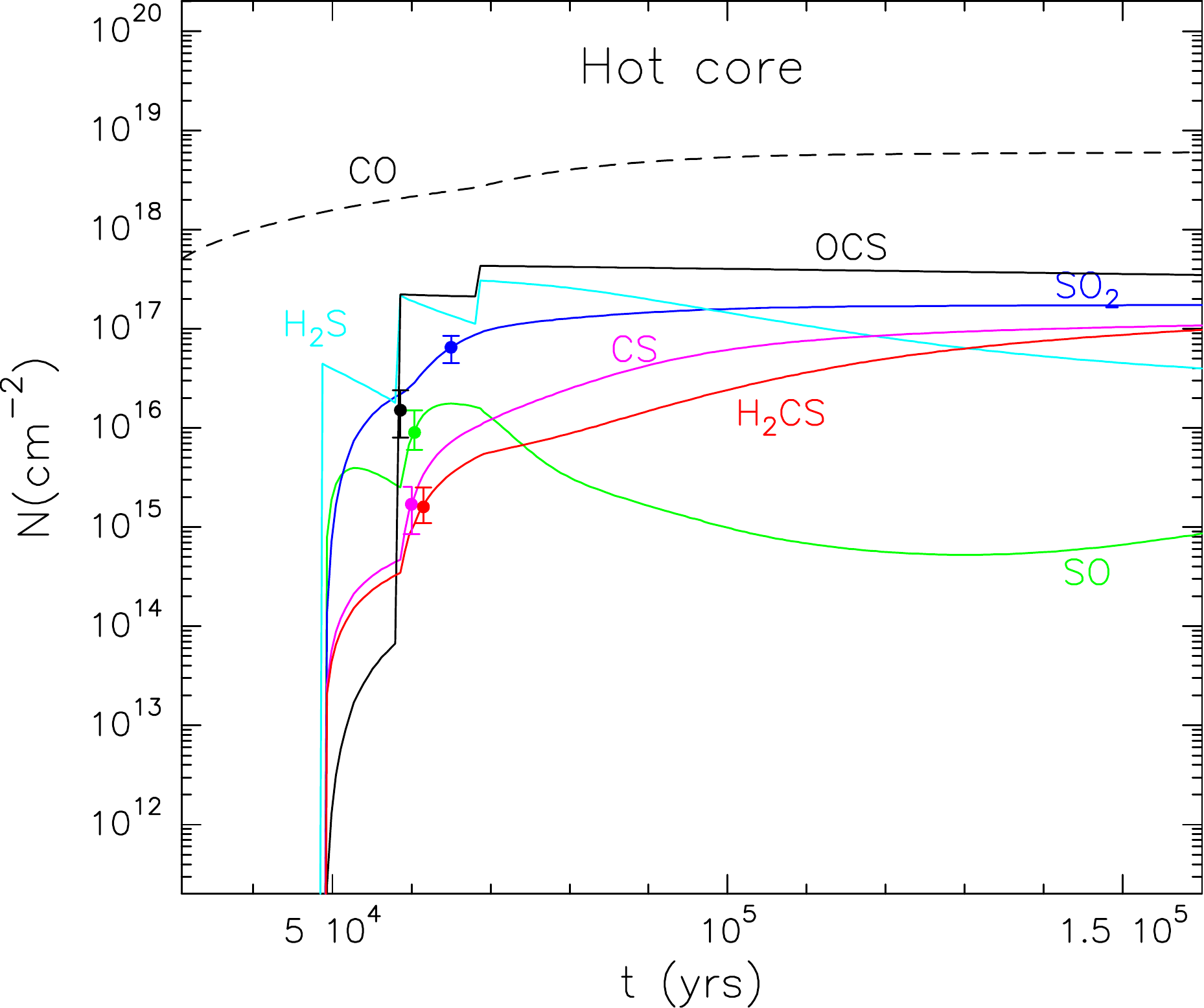}
   \caption{Sulphur-bearing species column densities as a time function (gas phase) for the hot core Model 5 with a star mass of 10M$_{\odot}$, $f_r$=0.85, and 0.1S$_{\odot}$ sulphur abundance. The points indicate the observational results of SO, SO$_{2}$, CS, OCS, H$_{2}$CS, and H$_{2}$S. In this case, S$^{+}$ freezes out to form OCS (50\%) and H$_{2}$S (50\%) on the mantles at the end of Phase I.}
   \label{figure:hot_core_model_2a}
   \end{figure} 

In Sect. \ref{hot core} we found that large variations in the percentage of S$^+$ ions that freeze out to form H$_2$S and OCS at the end of Phase I mainly affect the evolution of OCS, while the changes in the SO, SO$_2$, H$_2$S abundances are lower than one order of magnitude. 
From Fig. \ref{figure:efecto_porcentaje_S+}, we find that the more H$_2$S is formed in Phase I, the lower the OCS column densities in Phase II and the more similar to the observed value in the hot core, $N$(OCS)=1.5$\times$10$^{16}$ cm$^{-2}$. For this reason, the models listed in Table \ref{table:hotcore_model} have been obtained assuming that S$^{+}$ ions freeze out to form mainly ($\sim$95$\%$) H$_{2}$S at the end of Phase I. We also ran Model 5 assuming that 50$\%$ of S$^{+}$ in Phase I forms H$_2$S and the other 50$\%$ forms OCS (see Fig. \ref{figure:hot_core_model_2a}): we find that while the chemical evolution of CS, SO, SO$_2$, and H$_2$CS barely changes with respect to Fig. \ref{figure:hotcore_model_5}, as expected, the values for OCS present a difference of one order of magnitude, although it is still possible to reproduce the observations for this species. In the case of H$_2$S, we do not reproduce its observational column density with the new assumptions, although the difference between the model results shown in Fig. \ref{figure:hot_core_model_2a} and the H$_2$S observations is lower than one order of magnitude.

\subsection{Plateau}
\label{comparison_plateau}

We ran eight plateau models (Table \ref{table:plateau_model}), varying the input parameters of initial sulphur abundance, the maximum shocked gas temperature ($T$$^{\mathrm{max}}_{\mathrm{shock}}$), and the final gas density. Figures showing these models correspond to the Phase II of the plateau where $t$=0 indicates the start of the non-dissociative shock action leading to a fast increase in the gas temperature from $T$=10 K up to 1000 and 2000 K (depending on the model). After the passage of the shock, the gas rapidly (within hundred of years) cools down to $\sim$80 K following the temperature profiles shown in Fig. \ref{figure:shock_temperature}.

\begin{table}
\caption{Plateau chemical models.}             
\centering   
\begin{tabular}{c c c c }     
\hline\hline    
Model & Sulphur abundance & $T$$^{\mathrm{max}}_{\mathrm{shock}}$ & Density  $n$$\mathrm{_H}$         \\ 
     & (S$_{\odot}$) & (K)                                        &  (cm$^{-3}$)  \\
\hline 

   1 & 0.01 & 2000 & 5$\times$10$^{5}$ \\
   2 & 0.1  & 2000 & 5$\times$10$^{5}$ \\  
   3 & 0.01 & 2000 & 5$\times$10$^{6}$ \\
   4 & 0.1  & 2000 & 5$\times$10$^{6}$ \\
\hline 
   5 & 0.01 & 1000 & 5$\times$10$^{5}$ \\
   6 & 0.1  & 1000 & 5$\times$10$^{5}$ \\
   7 & 0.01 & 1000 & 5$\times$10$^{6}$ \\
   8 & 0.1  & 1000 & 5$\times$10$^{6}$ \\
\hline
\label{table:plateau_model}                 
\end{tabular}
\tablefoot{Column 1 indicates the model, Col.2 the initial sulphur abundance in solar units, Col.3 the maximum shocked gas temperature, and Col. 4 the gas density at Phase II. \\
}
\end{table}

\begin{figure}
   \centering 
   \includegraphics[angle=0,width=8cm]{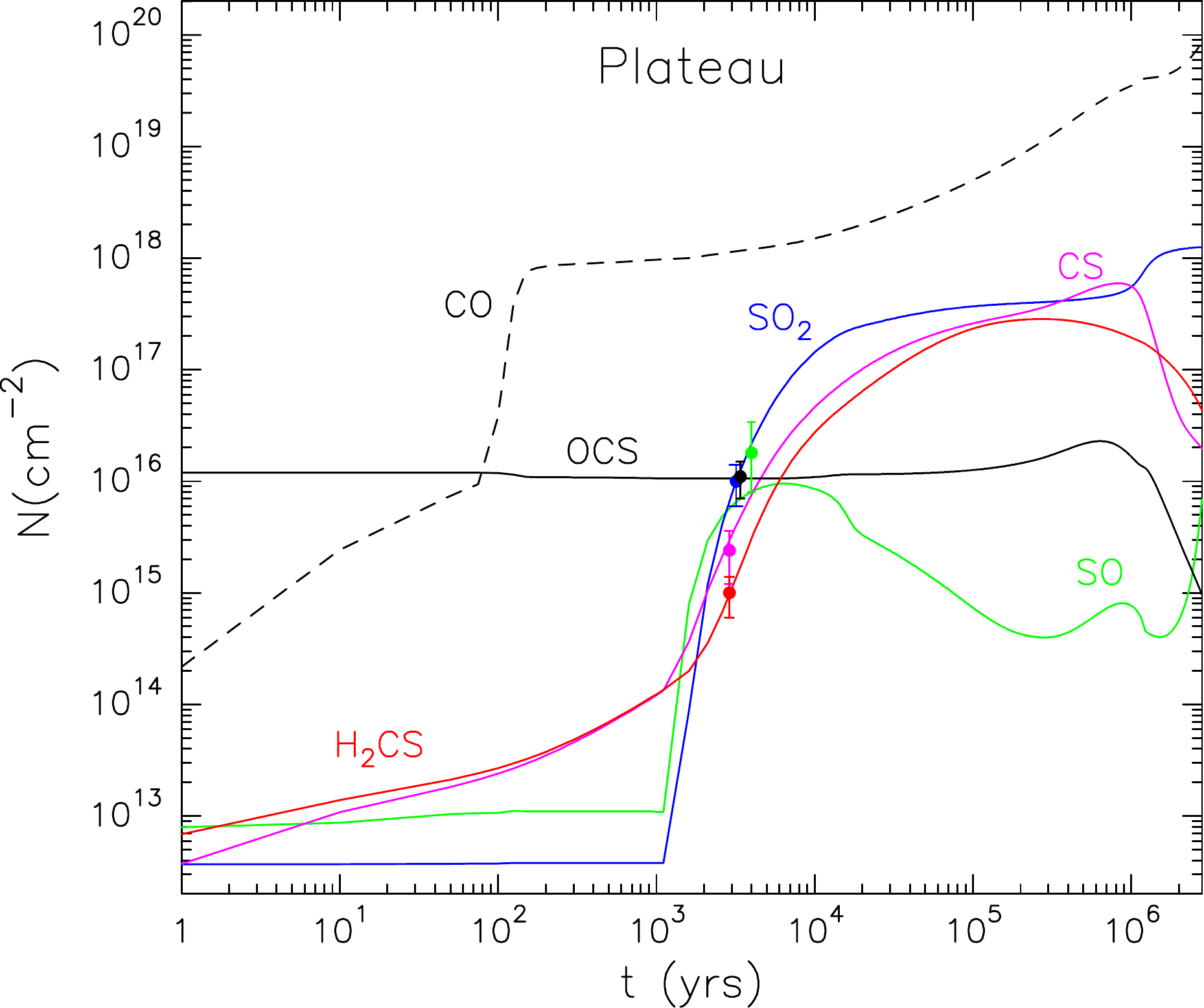}
   \caption{Sulphur-species column densities as a time function at Phase II for the plateau Model 4 with a final gas density of 5$\times$10$^{6}$ cm$^{-3}$, an initial sulphur abundance of 0.1S$_{\odot}$, and $T$$^{\mathrm{max}}_{\mathrm{shock}}$=2000 K.}
   \label{figure:plateau_model_17}
   \end{figure}

Figures \ref{figure:plateau_model_17} and \ref{figure:plateau_model_21} (Appendix \ref{Figures of Plateau}) show the column densities of sulphur-bearing molecules obtained as a function of time from Models 4 and 8. In both cases, the hydrogen density is  $n$$\mathrm{_H}$=5$\times$10$^{6}$ cm$^{-3}$ (recall that this is the final density at the end of Phase I), the initial sulphur abundance 0.1S$_{\odot}$, and only the maximum shocked gas temperature is varied. Taking the error bars for the observed column densities of each species into account, we find in both cases a time at which all species are matched well to the observations.

\begin{figure}
   \centering 
   \includegraphics[angle=0,width=8cm]{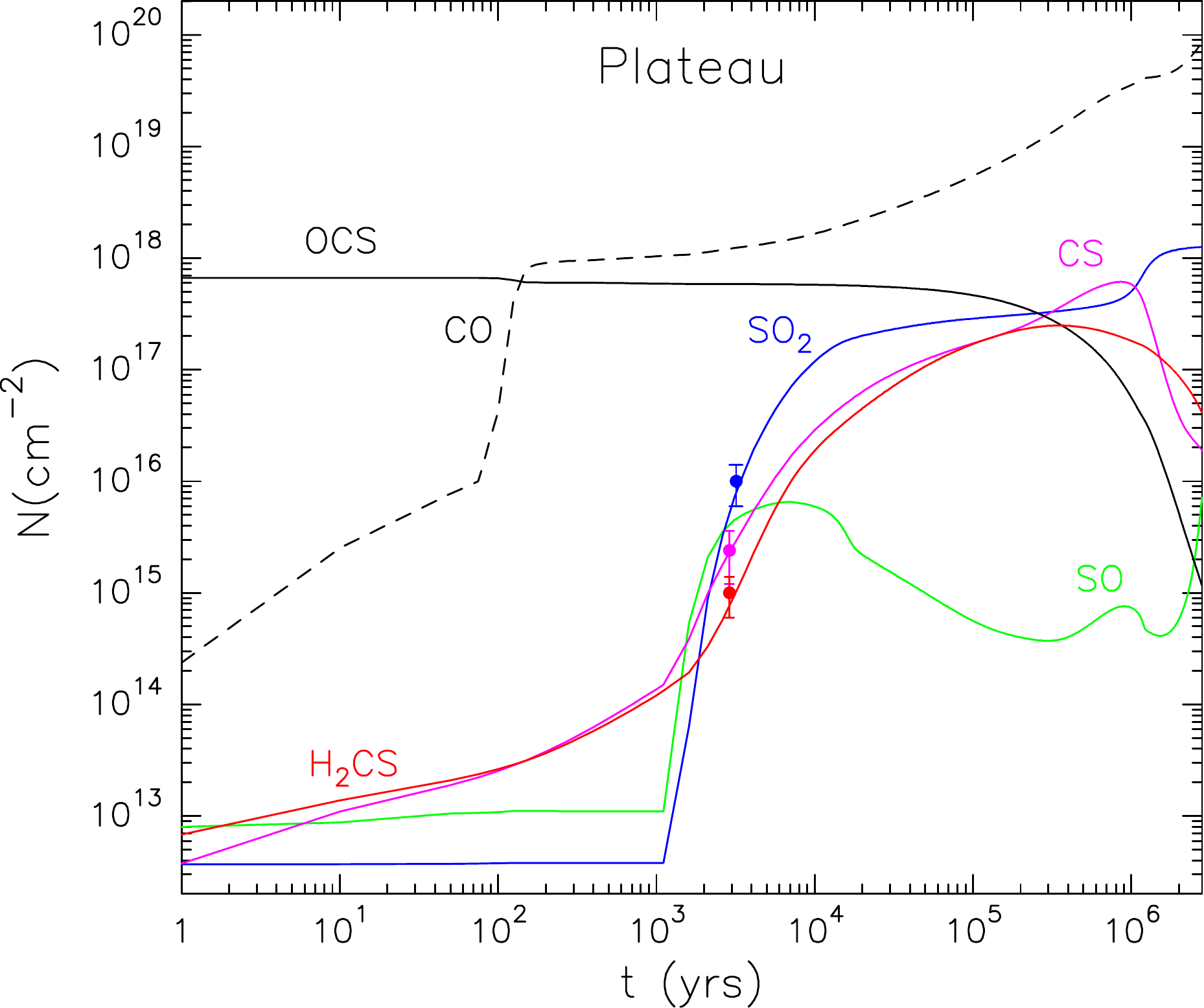}
   \caption{Sulphur-species column densities as a time function at Phase II for a plateau model with a density of 5$\times$10$^{6}$ cm$^{-3}$, an initial sulphur abundance of 0.1S$_{\odot}$, and $T$$_{\mathrm{shock}}$=2000 K. In this model, S$^{+}$ freezes out to form OCS (50$\%$) and H$_{2}$S (50$\%$) on the mantles at the end of Phase I.}
   \label{figure:plateau_model_1a}
   \end{figure} 

If we decrease the hydrogen density (constant along all Phase II) by one order of magnitude, i.e. to 5$\times$10$^{5}$ cm$^{-3}$, (see Figs. \ref{figure:plateau_model_19} and \ref{figure:plateau_model_23} showing Models 2 and 6, respectively), the maximum SO column density is to up to one order of magnitude lower than the observed one. The same is found for OCS in the case where $T$$^{\mathrm{max}}_{\mathrm{shock}}$=1000 K (Fig. \ref{figure:plateau_model_23}), while for Model 19 (Fig. \ref{figure:plateau_model_19}) with $T$$^{\mathrm{max}}_{\mathrm{shock}}$=2000 K the difference between the model and the observed value of OCS is less than one order of magnitude.  On the other hand, if we consider an initial sulphur abundance of 0.01S$_{\odot}$ (Model 3, Fig. \ref{figure:plateau_model_18}), we also obtain values for SO and OCS up to one order of magnitude less than the observed ones. 
This suggests that models of plateau assuming an initial sulphur abundance of 0.1S$_{\odot}$, as well as a final density of 5$\times$10$^{6}$ cm$^{-3}$ (Models 4 and 8), provide a much better fit to the observed column densities of the sulphur-bearing species. Models 4 and 8 only differ in the maximum temperature reached by the shocked gas (2000 and 1000 K, respectively). Our derived maximum shock temperatures agree with previous works: Maret et al. (2001) found a temperature of the shocked gas in Orion KL of 1500 K (assuming an angular extent of 40$\arcsec$) and Sempere et al. (2000) found $T$$^{\mathrm{max}}_{\mathrm{shock}}$=1500-2000 K using parameters representative of the H$_2$ peak 1 region with 10$\arcsec$ of size.  
This may suggest that the maximum shocked gas temperature reached in the plateau of Orion KL is closer to 2000 K than to 1000 K, therefore favouring Model 4 (Fig. \ref{figure:plateau_model_17}) in our present grid.
In this case, the observations are reproduced at $t$$\sim$2$\times$10$^3$ years. The temperature profile (Fig. \ref{figure:shock_temperature}) for a plateau model with a maximum shocked gas temperature of 2000 K indicates that at that time, the gas temperature is $\sim$90 K, which is consistent with the value considered in MADEX ($T$$_{\mathrm{k}}$=125 K) when the 20$\%$ of uncertainty assumed to model the observed line profiles of the sulphur-bearing species is taken into account.

As in the hot core case, the models listed in Table \ref{table:plateau_model} have also been run assuming that S$^{+}$ ions freeze out to form mainly ($\sim$95$\%$) H$_{2}$S at the end of Phase I and only $\sim$5$\%$ of OCS. Only for Model 4, where all species are well matched to the observations, we have also assumed that 50$\%$ of S$^{+}$ in Phase I forms H$_2$S and the other 50$\%$ forms OCS (see Fig. \ref{figure:plateau_model_1a}). In this case, we see that while the column densities of most sulphur-bearing species do not show any significant changes, the column densities of OCS are almost two orders of magnitude higher than those obtained in Fig. \ref{figure:plateau_model_17}. These results suggest that at the end of Phase I, most of S$^{+}$ is frozen out as H$_2$S.

\section{Discussion}
\label{Discussion}

\subsection{Sulphur reservoir on the grains}
\label{sulphur_reservoirs}

Although we do not have a firm detection of H$_2$S in interstellar ices, it does not necessarily imply that this molecule does not form, since, chemically, it should easily hydrogenate on grains. The non detection of solid H$_2$S in the interstellar medium can be due to either a fast dissociation\footnote{See Grim $\&$ Greenberg 1987, Moore et al. 2007, and Garozzo et al. 2010 who explained the failure of its detection in the solid phase from irradiation experiments with 200 keV protons. They found that H$_2$S column density has a very drop off at the beginning of irradiation, until almost all the H$_2$S molecules are decomposed.} or may even be an observational bias because this ice is particularly difficult to observe. Geballe et al. (1985, 1991) inferred H$_2$S ice in the feature from 3.90 to 3.97 $\mu$m in W33A, a high-mass protostar; however, this detection is not fully supported in the literature. Van der Tak et al. (2003) argue that infrared observations do not support the assumption that H$_2$S is the main S reservoir in grain mantles, and they provide the ISO-SWS observations of W33A as an example. One of the problems for identifying the 3.925 $\mu$m (2548 cm$^{-1}$) band of H$_2$S in H$_2$O-rich ice mantles was the lack of laboratory spectra of H$_2$S embedded in an H$_2$O matrix, which is expected to affect this band significantly.

Mumma and Charnley (2011) present a list with the percentage of the observed abundances (and upper limits for the non-observed ices) of several species relative to water in ices of massive protostars. For H$_2$S the value is $<$3$\times$10$^{-3}$-10$^{-2}$ and for OCS is 4$\times$10$^{-4}$-2$\times$10$^{-3}$ (obtained by Palumbo et al. 2007 with data from a sample of eight dense molecular clouds). The H$_2$S and OCS abundances relative to H$_2$O that we obtain at the end of Phase I in the hot core Model 5 (Fig. \ref{figure:hotcore_model_5} where all the species are well matched to the observations considering that S$^+$ freezes out to form mainly H$_2$S) are 3$\times$10$^{-3}$ and 2$\times$10$^{-4}$ for H$_2$S and OCS, respectively (see Table \ref{table:abundancesH2S}). Although the OCS abundance relative to water obtained with the model is slightly lower, both results (for H$_2$S and OCS) are consistent with those listed in Mumma and Charnley (2011).

\begin{table}
\caption{Abundances (relative to H) of the species H$_2$S, OCS, and H$_2$O on the mantles of the dust grains at the end of Phase I.}             
\centering   
\begin{tabular}{l l l l l}     
\hline\hline 
Mantle species  m(x)  &  Abundance  \\
\hline   
m(H$_2$S)             & 1.357$\times$10$^{-6}$ \\  
m(OCS)                & 7.300$\times$10$^{-8}$ \\
m(H$_2$O)             & 4.278$\times$10$^{-4}$ \\
\hline
m(H$_2$S)/m(H$_2$O)   & 3$\times$10$^{-3}$    \\
m(OCS)/m(H$_2$O)      & 2$\times$10$^{-4}$  \\
\hline
\label{table:abundancesH2S}                 
\end{tabular}
\end{table}

\subsection{Sulphur chemical evolution in the hot core}
\label{sulphur_chemical_evolution_hot_core}

We now examine the chemistry of sulphur-bearing species in Model 5 in more detail.  
In Fig. \ref{figure:hotcore_model_5}, we observe how H$_{2}$S is evaporated at different evolutionary stages in Phase II, in agreement with Viti et al. (2004). Each evaporation is followed by a sharp decrease in its column density owing to dissociation processes. The evolution of the column densities of SO and SO$_2$ are closely related; whereas SO is predicted to be mainly formed by the reaction

\begin{equation}
\mathrm{O + SH \rightarrow SO + H}, 
\label{equation:OH}
\end{equation}

\noindent it is destroyed by reacting with OH, leading to the efficient formation of SO$_2$: 

\begin{equation}
\mathrm{OH + SO \rightarrow SO_2 + H}. 
\label{equation:OH}
\end{equation}

\noindent This sequence triggers a continuous increase in SO$_2$, whereas SO does not stop decreasing. As a result, the SO$_{2}$/SO ratio increases with time, in agreement with the results obtained by Wakelam et al. (2011), who modelled the sulphur chemistry of several massive dense cores. This is also consistent with Esplugues et al. (2013a), who concluded that SO$_{2}$ is a better tracer of warm gas than SO. For longer times, $t$$\gtrsim$8$\times$10$^{4}$ years since the star switches on, the increase in SO$_2$ starts to be limited as the reaction 

\begin{equation}
\mathrm{SO_2 + C \rightarrow SO + CO}, 
\label{equation:OH}
\end{equation}

\noindent becomes more efficient. This again produces SO and then column densities of both sulphur-bearing molecules stabilise.
On the other hand, we find that during the hot core evolution, the formation of CS and H$_2$CS is mainly the result of the reactions: 

\begin{equation}
\mathrm{CH_2 + S \rightarrow CS + H_2} 
\label{equation:OH}
\end{equation}

\noindent and

\begin{equation}
\mathrm{CH_3 + S \rightarrow H_2CS + H}, 
\label{equation:OH}
\end{equation} 

\noindent respectively. Although their abundances are low for a young hot core, these two molecules become, together with SO$_2$, the most abundant sulphur-bearing species for a more evolved hot core.

\subsection{Sulphur chemical evolution in the plateau}
\label{sulphur_chemical_evolution_plateau}

Analysing the evolution chemistry of the different species in Model 4, Fig. \ref{figure:plateau_model_17}, we see that for $t$$\textless$10$^3$ years, when the temperature of the region remains high ($T$$\gtrsim$1000 K) due to the shock effect, the column densities of OCS, SO, and SO$_{2}$ remain almost unchanged. In particular, during this period we find that the main formation routes of SO and SO$_2$ are through molecular oxygen:

\begin{equation}
\mathrm{O_2 + S \rightarrow SO + O} 
\label{equation:OH}
\end{equation}  

\noindent and 

\begin{equation}
\mathrm{O_2 + SO \rightarrow SO_2 + O}. 
\label{equation:OH}
\end{equation}

\noindent For $t$$\gtrsim$10$^3$ years, we find that the chemistry drastically changes, especially for SO and SO$_{2}$, which increase their column densities up to five orders of magnitude. In this period where the gas has been cooled, we find that SO is mainly formed through OH:

\begin{equation}
\mathrm{OH + S \rightarrow SO + H}. 
\label{equation:OH}
\end{equation}

\noindent OH is also the main one responsible for the destruction of SO and the formation of SO$_2$, through the reaction

\begin{equation}
\mathrm{OH + SO \rightarrow SO_2 + H}. 
\label{equation:OH}
\end{equation}

\begin{figure}[h!]
   \centering 
   \includegraphics[angle=-90,width=8cm]{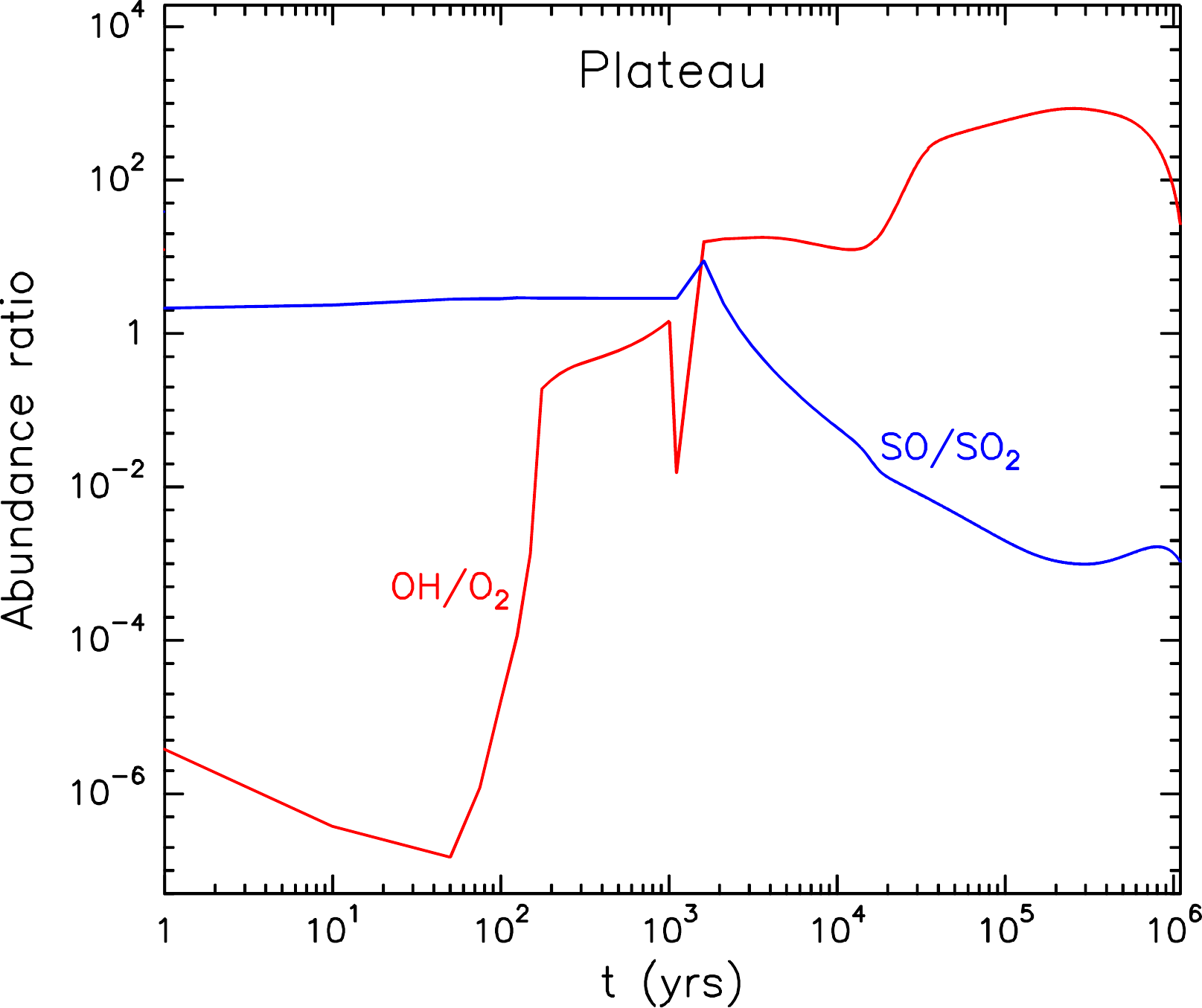}
   \caption{OH/O$_2$ and SO/SO$_2$ ratios along the time for the plateau warm phase (Phase II).}
   \label{figure:ratio_OH-O2}
   \end{figure}

\noindent With this reaction all sulphur monoxide is rapidly converted into SO$_2$, leading to a rapid decrease in the SO/SO$_2$ ratio with time.
Goicoechea et al. (2006b) found a fractional abundance of $X$(OH)=(0.5-1)$\times$10$^{-6}$ in the plateau and Goldsmith et al. (2011) obtained $X$(O$_2$)=(0.3-7)$\times$10$^{-6}$, deducing that this value is the result of an enhancement of $X$(O$_2$) by shocks. In fact, we have represented the evolution of the OH/O$_2$ ratio, Fig. \ref{figure:ratio_OH-O2}, and our model effectively predicts that during the shock action, the abundance of O$_2$ is much larger than that of OH, whereas we find the opposite for large times.
The abundances, as well as the dominant formation/destruction reactions of SO and SO$_2$ at different stages, indicate that a chemical transition (with similar abundances of O$_2$ and OH) is currently taking place in the plateau, where SO and SO$_2$ reactions change from being dominated by O$_2$ to being dominated by OH. In the case of CS and H$_{2}$CS (which are formed through the reaction of S with CH$_2$ and CH$_3$, respectively, for $t$$\lesssim$10$^5$ years), we observe that their column densities increase with time, and become, together with SO$_{2}$, the most abundant sulphur-molecules in an evolved plateau.

\section{Summary and conclusions}
\label{conclusions}

We have modelled the sulphur chemistry of the Orion KL cloud for two components, the hot core and the plateau shocked gas. We investigated a wide range of parameters, such as the gas density, the accretion efficiency, the initial sulphur abundance, the mass of the formed star, the shock temperature, and different chemical paths on the grains leading to different reservoirs of sulphur on the ice grain mantles.

We first analysed the effect of varying these parameters mainly on the time evolution of the SO and SO$_{2}$ abundances, during the warm-up phase. Our results can be summarized as follows:

\begin{enumerate}
\renewcommand{\labelenumi}{$\bullet$ }
\item {\it The mass of the central star}: its variation barely affects the SO and SO$_{2}$ abundances at early times of the warm up phase. For $t$$>$2$\times$10$^4$ years, the evolution times for both molecules present a difference of $\sim$5$\times$10$^4$ years between stars with 5M$_{\odot}$ and 15M$_{\odot}$.
\end{enumerate}

\begin{enumerate}
\renewcommand{\labelenumi}{$\bullet$ }
\item {\it Efficiency of gas accretion on the grains}: for an early hot core ($t$$<$5$\times$10$^4$ years), the accretion efficiency of species onto grain surfaces in the cold phase plays an important role in the abundances of SO and SO$_{2}$ present in the warm gas phase, leading to differences of more than two orders of magnitude. As the hot core evolves, these differences become negligible.
\end{enumerate}

\begin{enumerate}
\renewcommand{\labelenumi}{$\bullet$ }
\item {\it Maximum temperature of the shocked gas in the plateau}: from the time the shock starts and up to $\sim$10$^5$ years, as the maximum temperature reached increases, lower SO abundances and larger SO$_2$ abundances are obtained.
\end{enumerate}

\begin{enumerate}
\renewcommand{\labelenumi}{$\bullet$ }
\item {\it Depletion of S atoms versus S$^+$ ions}: varying the percentage of neutral sulphur that freezes on grains during the cold phase to form H$_{2}$S or OCS barely affects the sulphur reservoirs on the mantle. In contrast, varying the percentage of S$^{+}$ that freezes out significantly affects the OCS abundances. 
This suggests that most of the sulphur present in cold phase must have been in S$^{+}$ ions, consistent with an inhomogeneous and clumpy cloud where UV photons penetrate the gas, ionising S atoms. Our models indicate that S$^{+}$ ions depleted onto grains mainly form H$_{2}$S at the end of this phase. 
\end{enumerate}

Finally, we compared the results obtained from the models with observations of sulphur-bearing molecules (SO, SO$_{2}$, CS, OCS, H$_{2}$S, and H$_{2}$CS) in Orion KL.
It is possible to reproduce the observations of these species in the hot core by assuming models with an initial sulphur abundance of 0.1S$_{\odot}$ and a hydrogen density of 10$^7$ cm$^{-3}$, as well as with models assuming 0.01S$_{\odot}$ and $n$$_{\mathrm{H}}$=10$^8$ cm$^{-3}$, with a stellar mass higher than 5M$_{\odot}$ in both cases. By also considering HC$_3$N and DC$_3$N observations from a previous study, we find that high (85$\%$) accretion efficiency during the cold phase of the cloud is necessary, favouring Model 5.
For $t$$<$10$^5$ years, we find that the main SO formation route is through the reaction of SH with oxygen. SO$_2$ is formed by destroying SO when it reacts with OH, while CS and H$_2$CS, are mainly formed through CH$_2$ and CH$_3$, respectively.

For the plateau, it is possible to reproduce the observations of the sulphur-bearing molecules through models with an initial sulphur abundance of 0.1S$_{\odot}$ and a hydrogen density during the warm-phase of at least 5$\times$10$^{6}$ cm$^{-3}$. 
We deduce that since the start of the shock and for $\sim$10$^3$ years, the column densities of SO$_{2}$, SO, and OCS do not significantly change. For longer times, however, the chemistry drastically changes; the SO/SO$_2$ ratio rapidly decreases with time until the SO column densities become more than two orders of magnitude lower than those of SO$_2$, CS, and H$_2$CS. 
In this component, we also find that a chemical transition occurs, where SO and SO$_2$ reactions change from being dominated by O$_2$ to being dominated by OH.

\begin{acknowledgements}
We thank the anonymous referee for valuable comments that greatly improved the manuscript.
We thank the Spanish MINECO for funding support from grants CSD2009-00038, AYA2009-07304, and AYA2012-32032. G.B.E. is supported by a CSIC grant JAE PreDoc2009. J.R.G. is supported by a Ram\'on y Cajal research contract.

\end{acknowledgements}

{}

\begin{appendix}
\section{Figures of hot core}
\label{Figures of Hot core}

The figures of this Appendix show hot core models in Phase II. The points indicate the observational results of SO, SO$_{2}$, CS, OCS, H$_{2}$CS, and H$_{2}$S.

\clearpage

\begin{figure}
\begin{center}
   \includegraphics[angle=0,width=7.5cm]{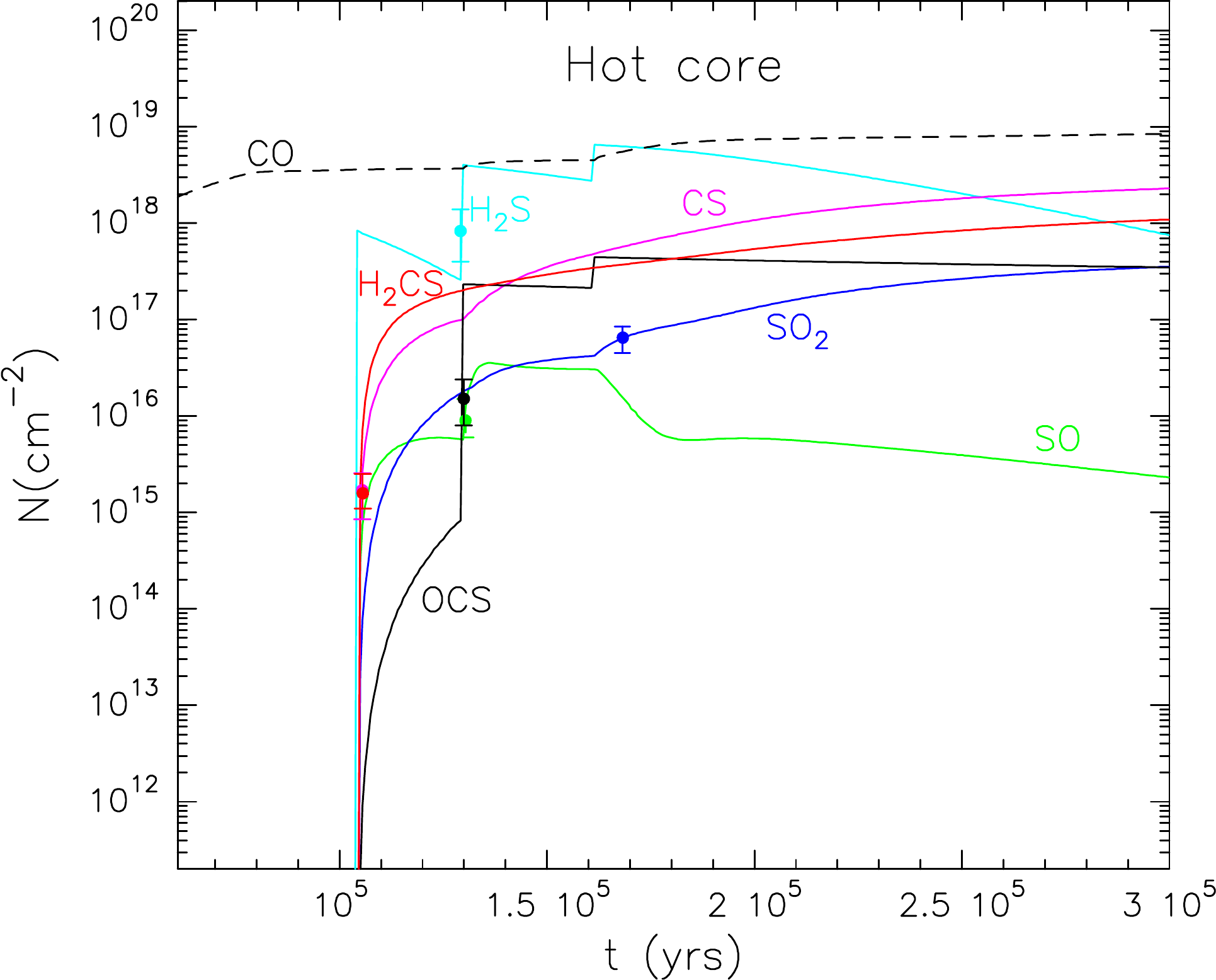} 
   \caption{Sulphur-bearing species column densities as a time function in a hot core model (gas phase) with a star mass of 5M$_{\odot}$, a solar sulphur abundance, 1S$_{\odot}$, and gas density of 10$^7$ cm$^{-3}$ (Model 18). }
   \label{figure:hotcore_model_18}
   \end{center}
   \end{figure}

\begin{figure}
\begin{center}
   \includegraphics[angle=0,width=7.5cm]{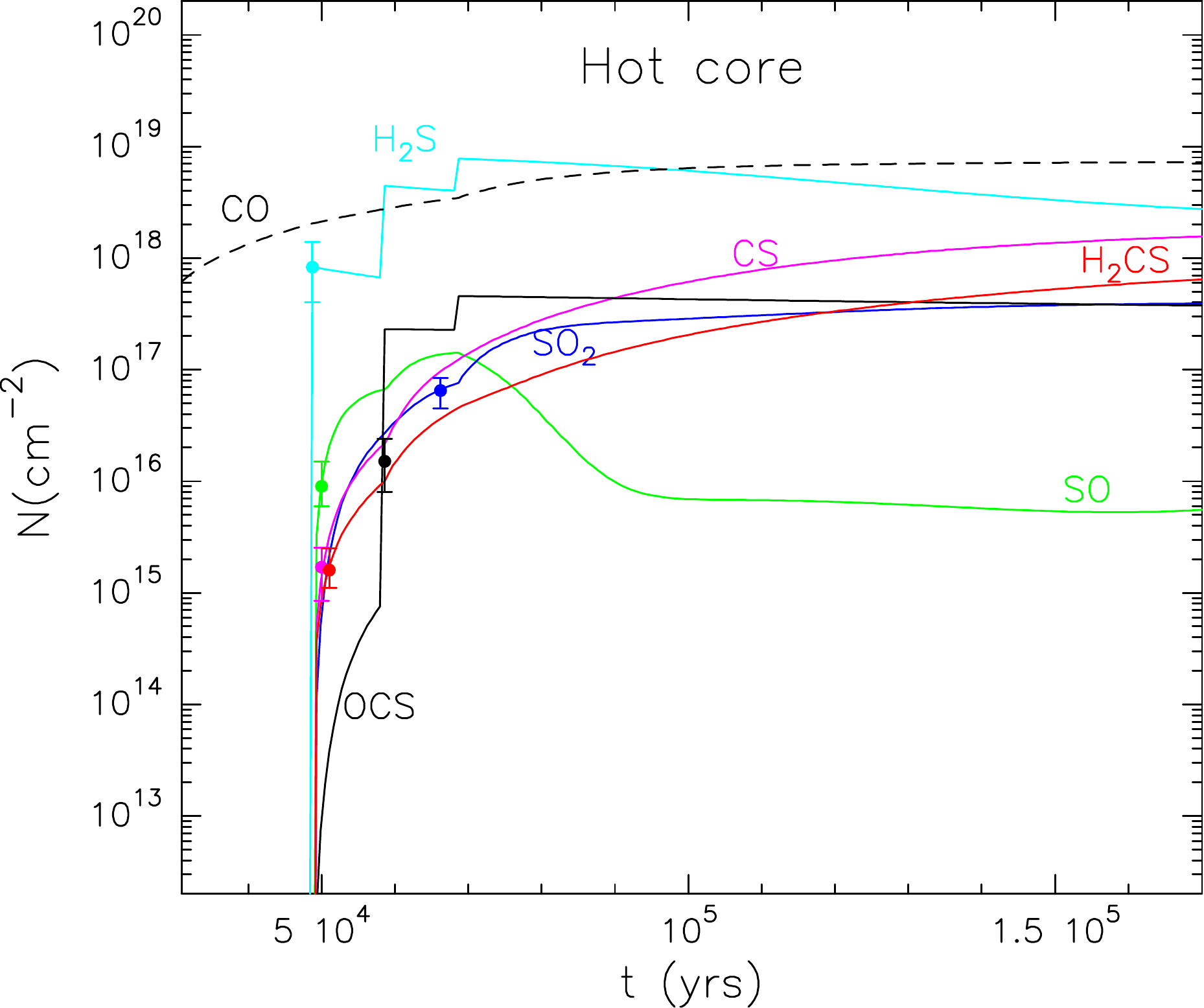} 
   \caption{Sulphur-bearing species column densities as a time function in a hot core model (gas phase) with a star mass of 10M$_{\odot}$, 1S$_{\odot}$, and density of 10$^7$ cm$^{-3}$ (Model 17).}
   \label{figure:hotcore_model_17}
   \end{center}
   \end{figure}

\begin{figure}
\begin{center}
   \includegraphics[angle=0,width=7.5cm]{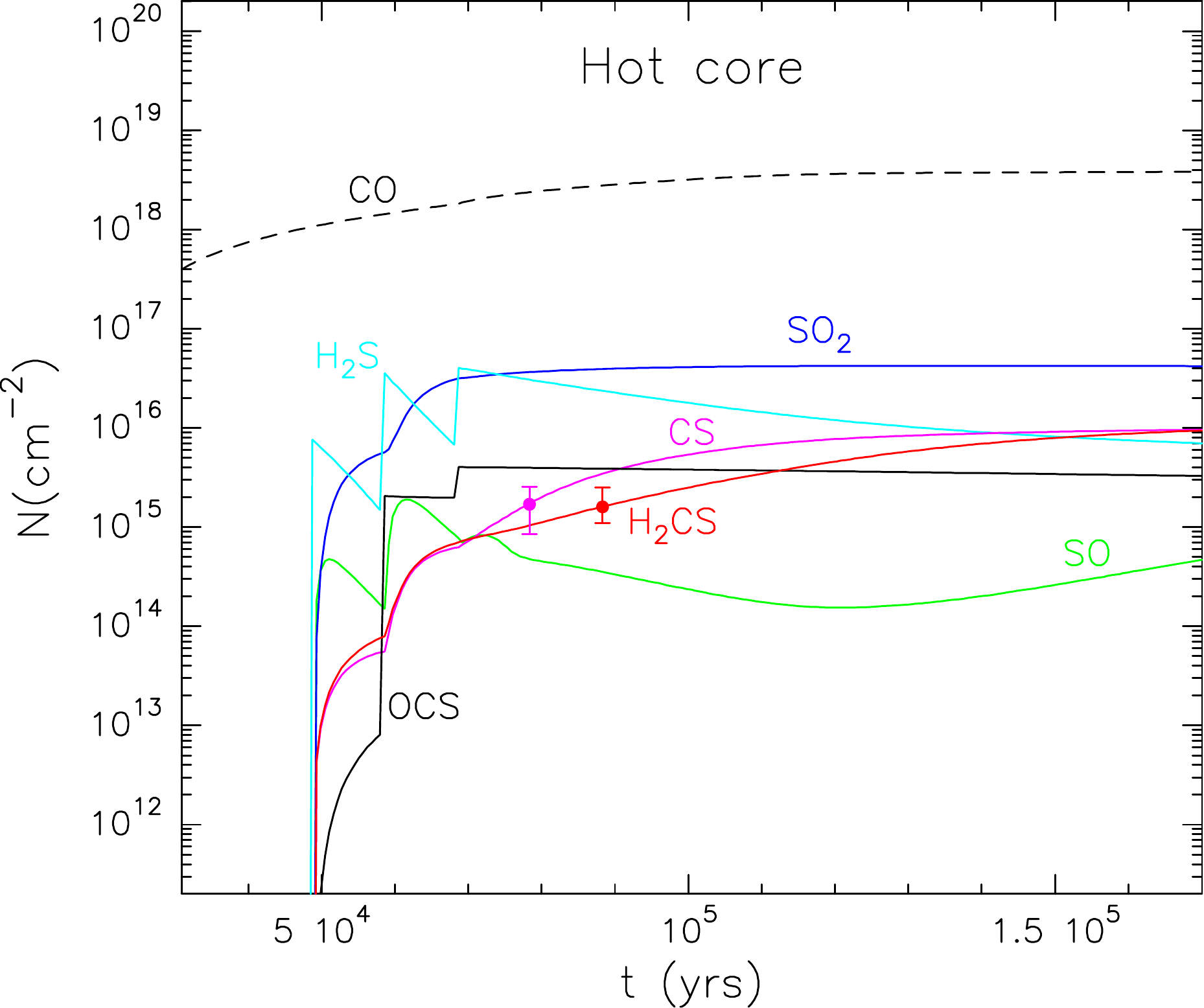} 
   \caption{Sulphur-bearing species column densities as a time function in a hot core model (gas phase) with a star mass of 10M$_{\odot}$, 0.01S$_{\odot}$, and density of 10$^7$ cm$^{-3}$ (Model 11).}
   \label{figure:hotcore_model_11}
   \end{center}
   \end{figure}

\begin{figure}
\begin{center}
   \includegraphics[angle=0,width=7.5cm]{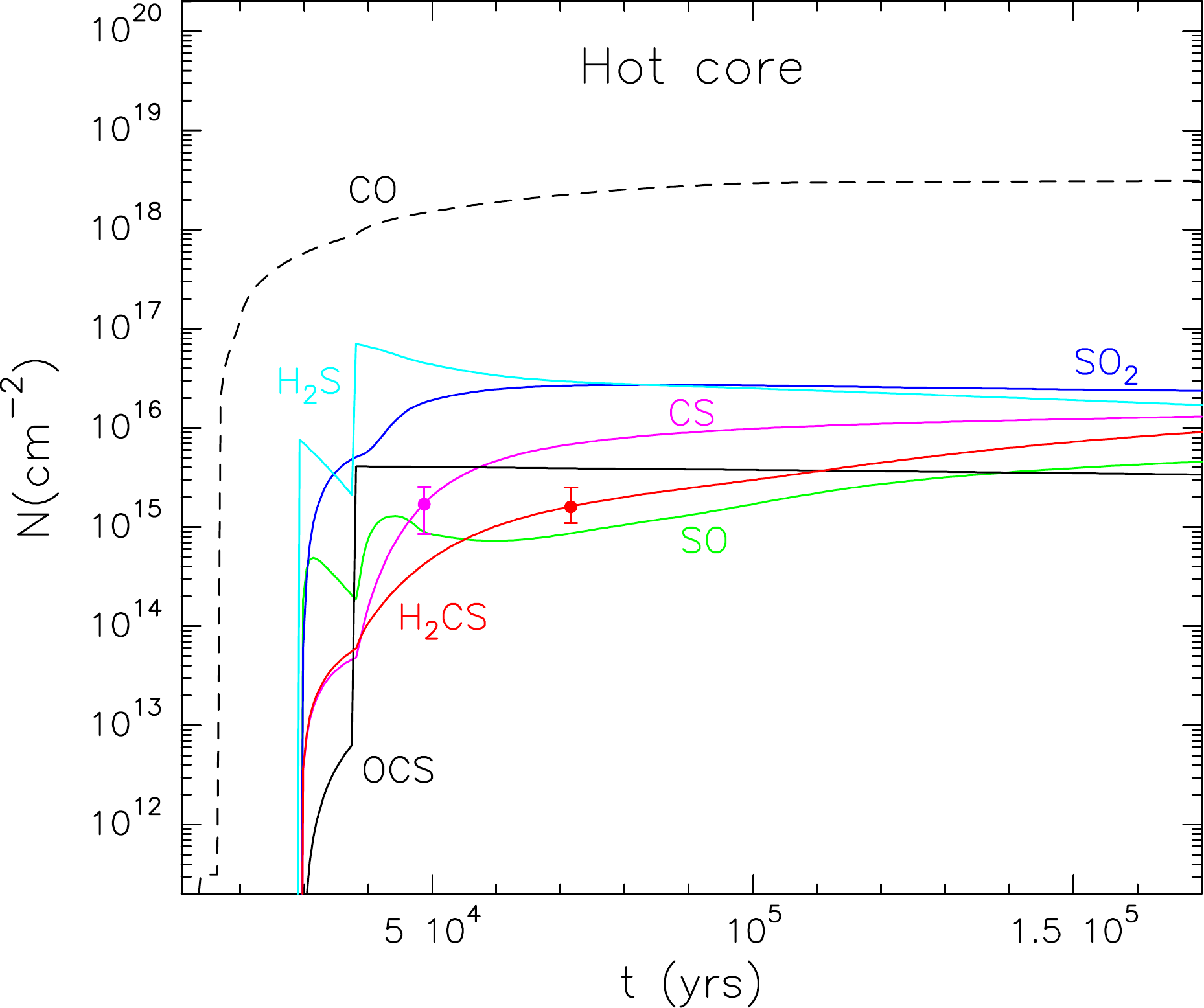} 
   \caption{Sulphur-bearing species column densities as a time function in a hot core model (gas phase) with a star mass of 15M$_{\odot}$, 0.01S$_{\odot}$, and density of 10$^7$ cm$^{-3}$ (Model 10). }
   \label{figure:hotcore_model_10}
   \end{center}
   \end{figure}

\begin{figure}
\begin{center}
   \includegraphics[angle=0,width=7.5cm]{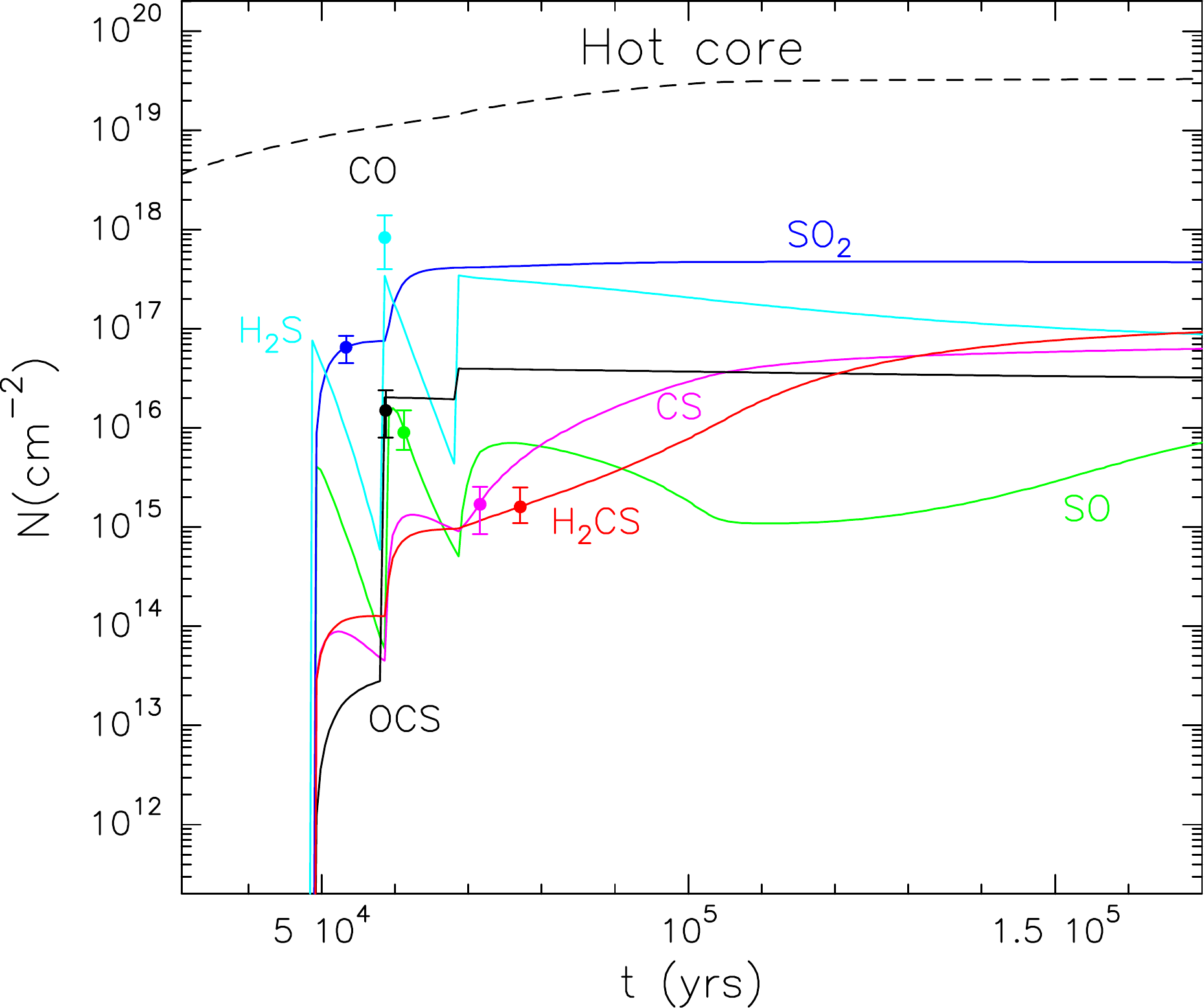} 
   \caption{Sulphur-bearing species column densities as a time function in a hot core model (gas phase) with a star mass of 10M$_{\odot}$, 0.01S$_{\odot}$, and a density of 10$^8$ cm$^{-3}$ (Model 11\textit {b}).}
   \label{figure:hotcore_model_11b}
   \end{center}
   \end{figure}

\begin{figure}
\begin{center}
\vspace{-0.3cm}
   \includegraphics[angle=0,width=7.5cm]{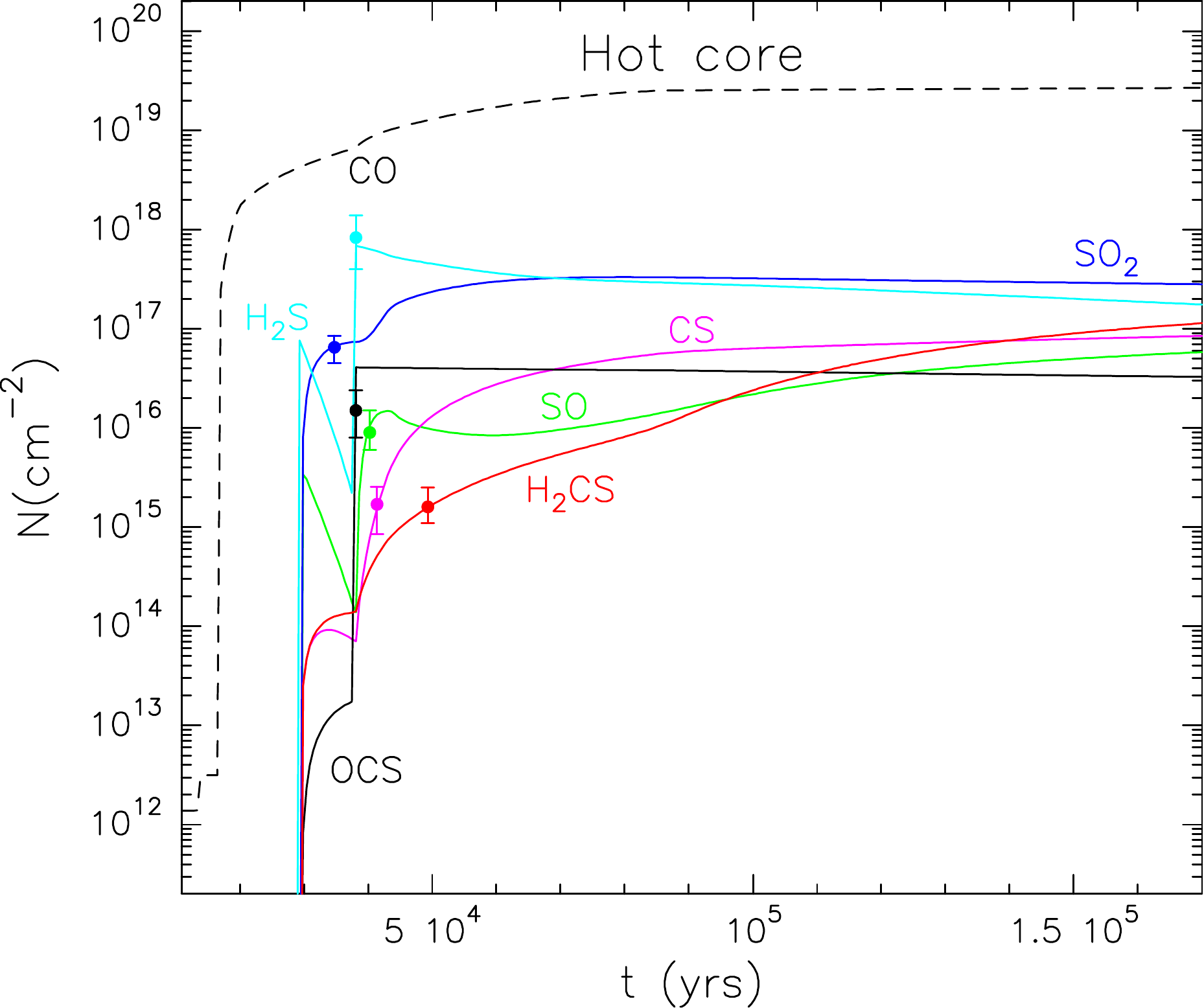} 
   \caption{Sulphur-bearing species column densities as a time function in a hot core model (gas phase) with a star mass of 15M$_{\odot}$, 0.01S$_{\odot}$, and a gas density of 10$^8$ cm$^{-3}$ (Model 10\textit {b}).}
   \label{figure:hotcore_model_10b}
   \end{center}
   \end{figure}

\begin{figure}
\begin{center}
   \includegraphics[angle=0,width=7.5cm]{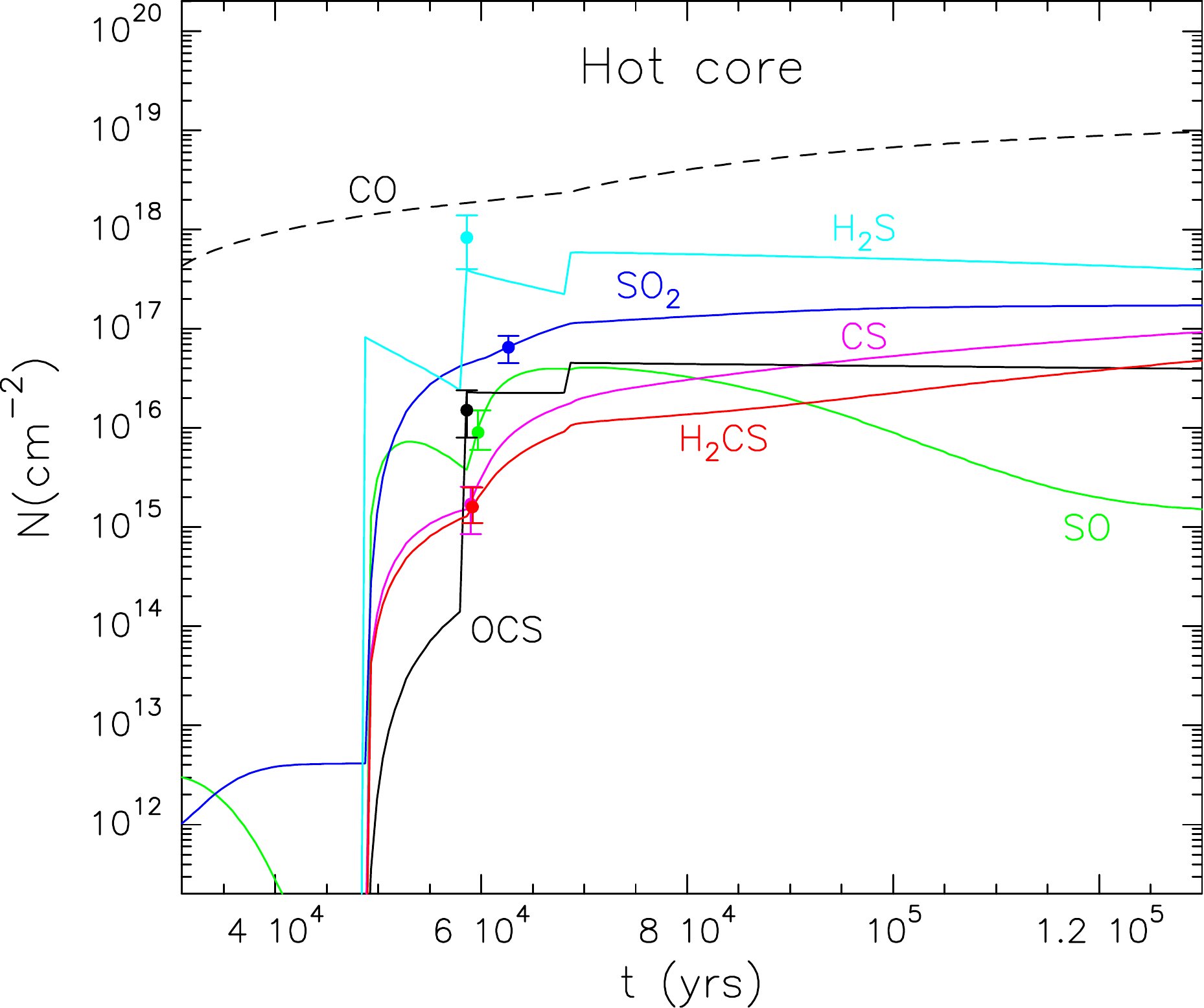} 
   \caption{Sulphur-bearing species column densities as a time function in a hot core model (gas phase) with a star mass of 10M$_{\odot}$, 0.1S$_{\odot}$, $f$$\mathrm{_r}$=0.3, and 10$^7$ cm$^{-3}$ of density (Model 2). }
   \label{figure:hotcore_model_2}
   \end{center}
   \end{figure}

\begin{figure}
\begin{center}
   \includegraphics[angle=0,width=7.5cm]{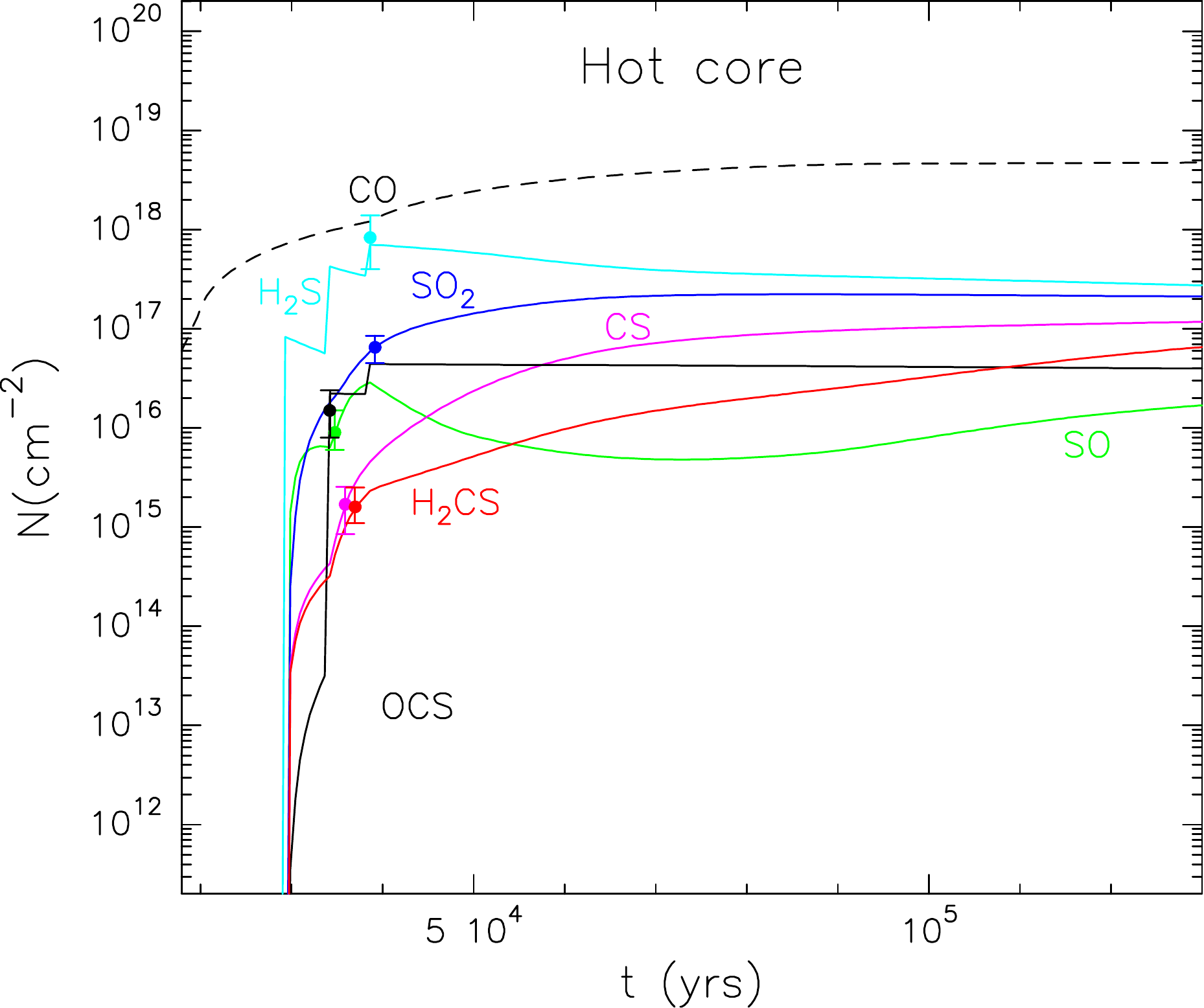} 
   \caption{Sulphur-bearing species column densities as a time function in a hot core model (gas phase) with a star mass of 15M$_{\odot}$, 0.1S$_{\odot}$, $f$$\mathrm{_r}$=0.85, and a density of 10$^7$ cm$^{-3}$ (Model 4).}
   \label{figure:hotcore_model_4}
   \end{center}
   \end{figure}

\clearpage

\section{Figures of plateau}
\label{Figures of Plateau}

The figures of this Appendix show plateau models in Phase II. The points indicate the observational results of SO, SO$_{2}$, CS, OCS, and H$_{2}$CS.

\begin{figure}[!hd]
   \centering 
   \includegraphics[angle=0,width=8.0cm]{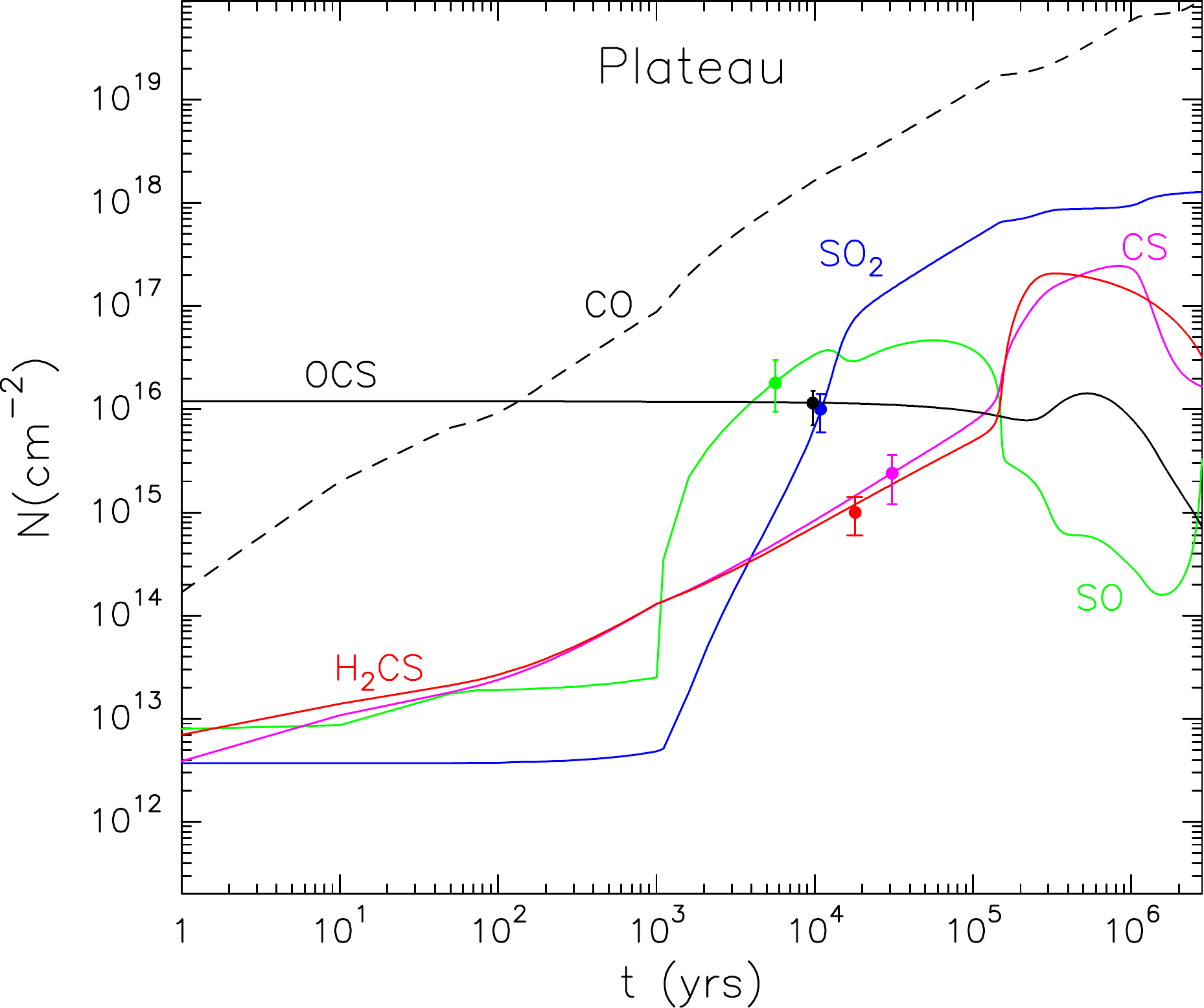}
   \caption{Sulphur-species column densities as a time function in the gas phase for a plateau model with a final gas density of 5$\times$10$^{6}$ cm$^{-3}$, an initial sulphur abundance of 0.1S$_{\odot}$, and $T$$^{\mathrm{max}}_{\mathrm{shock}}$=1000 K (Model 8).}
   \label{figure:plateau_model_21}
   \end{figure} 

\begin{figure}[!hd]
   \centering 
   \includegraphics[angle=0,width=8.0cm]{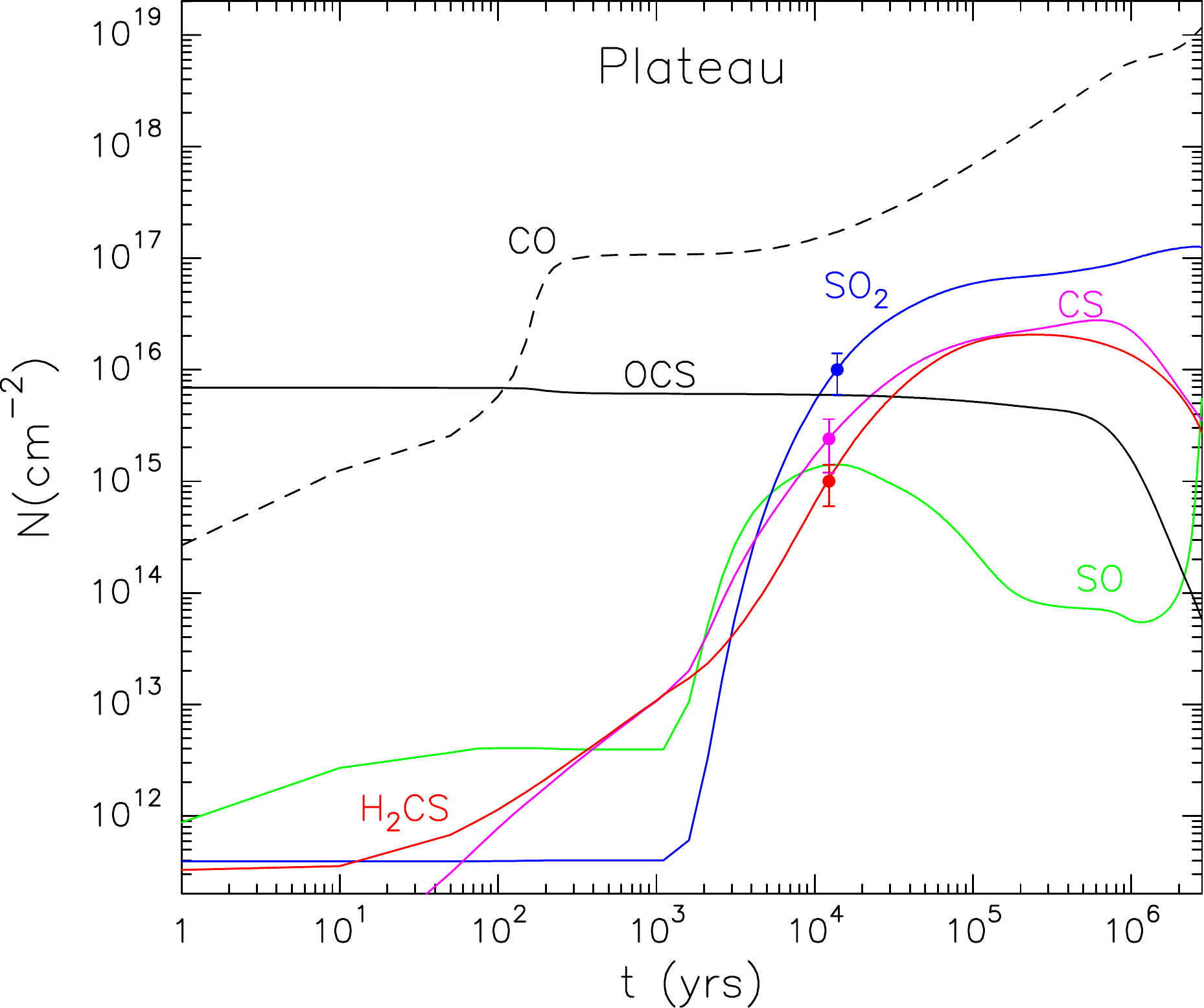}
   \caption{Sulphur-species column densities as a time function in the gas phase for a plateau model with a final gas density of 5$\times$10$^{5}$ cm$^{-3}$, an initial sulphur abundance of 0.1S$_{\odot}$, and $T$$^{\mathrm{max}}_{\mathrm{shock}}$=2000 K (Model 2).}
   \label{figure:plateau_model_19}
   \end{figure} 

\begin{figure}
   \centering 
   \vspace{2.4cm}
   \includegraphics[angle=0,width=8.0cm]{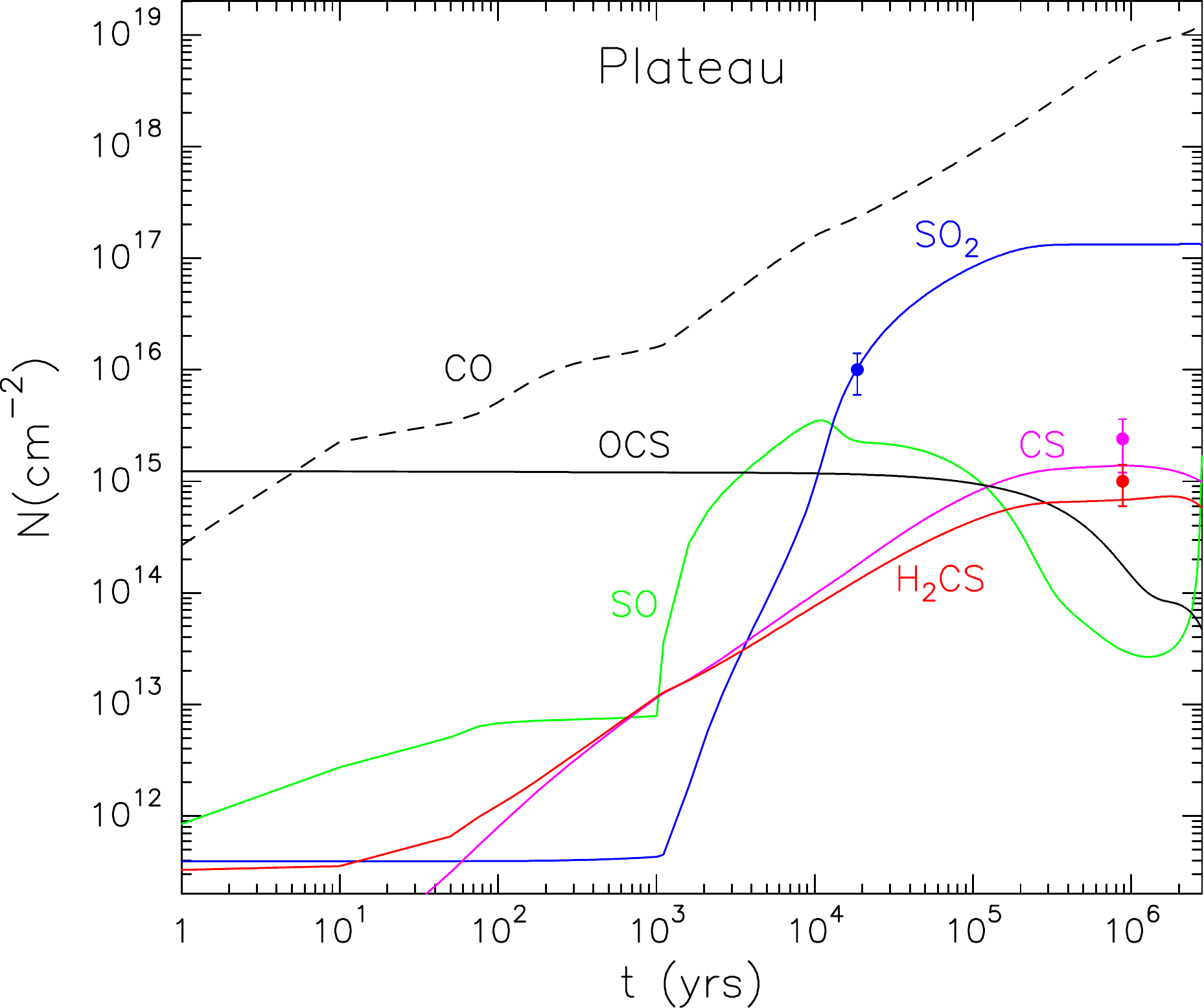}
   \caption{Sulphur-species column densities as a time function in the gas phase for a plateau model with a final hydrogen density of 5$\times$10$^{5}$ cm$^{-3}$, an initial sulphur abundance of 0.1S$_{\odot}$, and $T$$^{\mathrm{max}}_{\mathrm{shock}}$=1000 K (Model 6).}
   \label{figure:plateau_model_23}
   \end{figure} 

\begin{figure}
   \centering 
   \vspace{0.85cm}
   \includegraphics[angle=0,width=8cm]{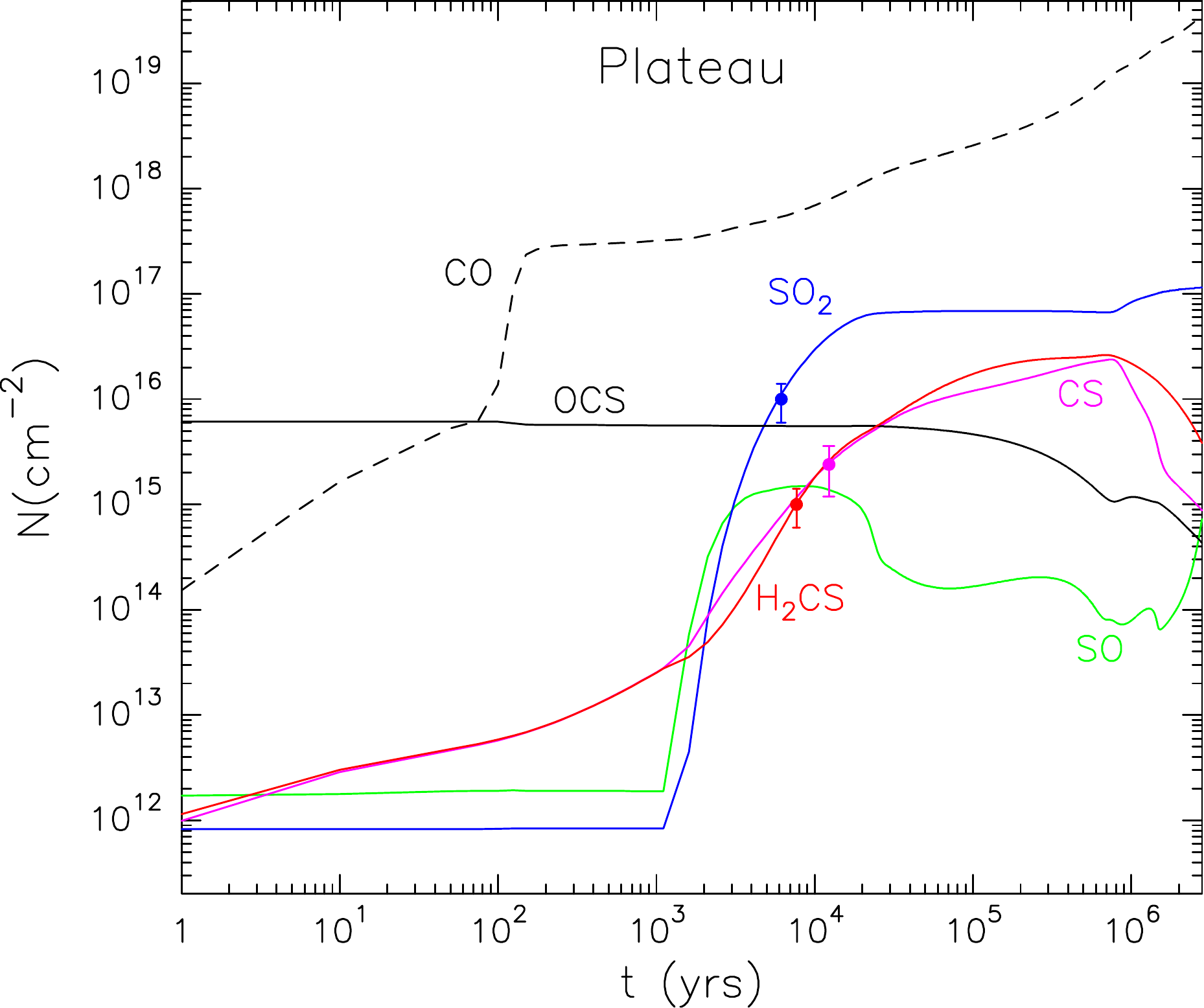}
   \caption{Sulphur-species column densities as a time function in the gas phase for a plateau model with a density of 5$\times$10$^{6}$ cm$^{-3}$, an initial sulphur abundance of 0.01S$_{\odot}$, and $T$$^{\mathrm{max}}_{\mathrm{shock}}$=2000 K (Model 3).}
   \label{figure:plateau_model_18}
   \end{figure} 


\end{appendix}


\begin{thebibliography}{}

   \bibitem[1989]{Anders89} Anders, E. \& Grevesse, N., 1989, GeCoA, 53, 197.
\bibitem[2005]{Asplund05} Asplund, M., 2005, ARA\&A, 43, 481.
   \bibitem[1997]{Bachiller97} Bachiller, R. \& P\'erez-Guti\'errez, M., 1997, ApJ, 487, L93-L96.
   \bibitem[2001]{Bachiller01} Bachiller, R., P\'erez-Guti\'errez, M., Kumar, M. S. N., \& Tafalla, M., 2001, A\&A, 372, 899.
   \bibitem[2005]{Bally05} Bally, J. \& Zinnecker, H., 2005, AJ, 129, 2281.   
   \bibitem[1989]{Bel89} Bel N., Lafen J.P.J., Viala Y. P., \& Loirelux E., 1989, A\&A, 208, 331.
    \bibitem[1998]{Bergin98} Bergin, E. A., Melnick, G. J., \& Neufeld, D. A., 1998, ApJ, 499, 777-792.
     \bibitem[2001]{Bergin01} Bergin, E. A., Ciardi, D. R., Lada, C. J., et al. 2001, ApJ, 557, 209-225.
    \bibitem[1987]{Blake87} Blake, G., Sutton, E. C., Masson, C. R., \& Plillips, T. G., 1987, ApJ, 315, 621.
     \bibitem[1988]{Brown88} Brown, P. D., Charnley, S. B., \& Millar, T. J., 1988, MNRAS, 231, 409.     
    \bibitem[1994]{Cernicharo94} Cernicharo, J., Gonz\'alez-Alfonso, E., Alcolea, J., et al. 1994, ApJ, 432, L59-L62. 
     \bibitem[2012]{Cernicharo12} Cernicharo, J., 2012, in ECLA-2011: Proceedings of the Europan Conference on Laboratory Astrophysics. Europan Astonomical Society Publications Series, 2012.
      \bibitem[1997]{Charnley97} Charnley, S. B., 1997, ApJ, 481, 396-405.     
      \bibitem[2004]{Collings2004} Collings, M. P., Anderson, M. A., Chen, R., et al. 2004, MNRAS, 354, 1133C.
       \bibitem[2014]{Crockettt14} Crockettt, N. R., Bergin, E. A., Neill, J. L., et al. 2014a, ApJ, 781, 114.      
       \bibitem[2014]{Crockettt14} Crockettt, N. R., Bergin, E. A., Neill, J. L., et al. 2014b, ApJ 787, 112.
      \bibitem[2012]{Druard12} Druard, C. \& Wakelam, V., 2012, MNRAS, 426, 354-359.       
      \bibitem[2013]{Esplugues13} Esplugues, G. B., Tercero, B., Cernicharo, J., et al. 2013a, A\&A, 556, A143.
         \bibitem[2013]{Esplugues13} Esplugues, G. B., Cernicharo, J., Viti, S., et al. 2013b, A\&A, 559, A51. 
    \bibitem[1975]{Gail75} Gail H. P. \& Sedlmayr E., 1975, A\&A, 41, 359.
     \bibitem[2010]{Garozzo10} Garozzo, M., Fulvio, D., Kanuchova, Z., et al. 2010, A\&A, 509, A67.    
     \bibitem[1985]{Geballe85} Geballe, T. R., Baas, F., Greenberg, J. M., \& Schutte, W., 1985, A\&A, 146, L6-L8.
    \bibitem[1991]{Geballe91} Geballe, T. R., 1991, MNRAS, 251, 24p-25p.
     \bibitem[1989]{Genzel89} Genzel, R. \& Stutzki, J., 1989, ARA\&A 27, 41-85.
    \bibitem[2006]{Goicoechea06} Goicoechea, J. R., Pety, J., Gerin, M., et al. 2006, A\&A, 456, 565-580.
     \bibitem[2006]{Goicoechea06} Goicoechea, J. R., Cernicharo, J., Lerate, M., et al. 2006b, ApJ, 641, L49-L52.
    \bibitem[2011]{Goldsmith11} Goldsmith, P. F., Liseau, R., Bell, T. A., et al. 2011, ApJ, 737, 96.
    \bibitem[1987]{Grim87} Grim, R. J. A. \& Greenberg, J. M., 1987, A\&A, 181, 168.
   \bibitem[1998]{Hatchell98} Hatchell, J., Thompson, M. A., Millar, T. J., \& MacDonald, G. H., 1998, A\&A, 338, 713.
    \bibitem[1998]{Harwit98} Harwit, M., Neufeld, D. A., Melnick, G. J., \& Kaufman, M. J., 1998, ApJ, 497, 105.
    \bibitem[1996]{Kaufman96} Kaufman, M. J. \& Neufeld, D. A., 1996, ApJ, 456, 611.
      \bibitem[2001]{Maret01} Maret, S., Caux, E., Baluteau, J. P., et al. 2001, ESASP, 460, 455.   
    \bibitem[1980]{McKee80} McKee, C. F.\& Hollenbach, 1980, ARA\&A, 18, 219.
      \bibitem[2000]{Molinari00} Molinari, S., Brand, J., Cesaroni, R., \& Palla, F., 2000, A\&A, 355, 617-628.
     \bibitem[2007]{Moore07} Moore, M. H., Hudson, R. L., \& Carlson, R. W., 2007, Icarus, 189, 409.       
     \bibitem[2011]{Mumma11} Mumma, M. J. \& Charnley, S. B., 2011, ARA\&A, 49, 471-524.
     \bibitem[1990]{Nejad90} Nejad, L. A. M., Williams, D. A. \& Charnley, S. B., 1990, MNRAS, 246, 183.      
     \bibitem[2001]{Neufeld01} Neufeld, D. A., 2001, ASPC, 235, 59.
      \bibitem[2011]{Oberg11} \"Oberg, K. I., Boogert, A. C. A., Pontoppidan, K. M., et al. 2011, IAUS Symposium 280.         
      \bibitem[2005]{Pagani05} Pagani, L., Pardo, J. R., Apponi, A. J., et al. 2005, A\&A, 429, 181-192.
       \bibitem[1997]{Palumbo97} Palumbo, M. E., Geballe, T. R., \& Tielens, A. G. G., 1997, ApJ, 479, 839-844.
    \bibitem[1978]{Pankonin78} Pankonin, V. \& Walmsley, C. M., 1978, A\&A, 64, 333-340.
       \bibitem[1987]{Plambeck87} Plambeck, R. L. \& Wright, M. C. H., 1987, ApJ, 317, L101-L105. 
    \bibitem[1992]{Rawlings92} Rawlings, J. M. C., Hartquist, T. W., Menten, K. M., \& Williams, D. A., 1992, MNRAS, 255.         
    \bibitem[2007]{Roberts07} Roberts, F. J., Rawlings, J. M. C., Viti, S., \& Williams, D. A., 2007, MNRAS, 733-742.       
    \bibitem[1999]{Ruffle99} Ruffle, D. P., Hartquist, T. W., Caselli, P., \& Williams, D. A., 1999, MNRAS, 306, 691.
    \bibitem[2000]{Sempere00} Sempere, M. J., Cernicharo, J., Lefloch, B., \& Gonz\'alez-Alfonso, E., 2000, ApJ, 530, L123-L127. 
    \bibitem[1978]{Spitzer78} Spitzer, L., 1978, {\it Physical processes in the interstellar medium}.
    \bibitem[2010]{Tercero10} Tercero, B., Cernicharo, J., Pardo, J. R., \& Goicoechea, J. R., 2010, A\&A, 517, A96.
    \bibitem[2003]{van der Tak03} van der Tak, F. F. S., Boonman, A. M. S., Braakman, R. \& van Dishoeck, E. F., et al. 2003, A\&A, 412, 133.    
    \bibitem[2007]{van der Tak07} van der Tak, F. F. S., Black, J. H., Sch\"oier, F. L., et al. 2007, A\&A, 468, 627.
   \bibitem[2001]{Viti01} Viti S., Caselli P., Hartquist T. W., \& Williams D. A., 2001, A\&A, 370, 1017. 
   \bibitem[2004]{Viti04} Viti, S., Collings, M.P, Dever, J.W., et al. 2004, MNRAS, 354, 1141.
      \bibitem[2011]{Wakelam11} Wakelam, V., Hersant, F., \& Herpin, F., 2011, A\&A, 529, 112.   
      \bibitem[1987]{Walmsley87} Walmsley, C. M., Hermsen, W., Henkel, C., et al. 1987, A\&A, 172, 311-315.
   \bibitem[1985]{Watt85} Watt, G. D., Millar, T. J., White, G. J., \& Harten, R. H., 1985, ESOC, 22, 381.
   \bibitem[1986]{Watt86} Watt, G. D., Millar, T. J., Glenn, J. W., \& Harten, R. H., 1986, A\&A, 155, 339-343.
      \bibitem[1988]{Welch88} Welch, W. J., 1988, ApL\&C, 26, 181-190.









\end{thebibliography}
\end{document}